\documentclass[a4paper,nofootinbib,20pt,showpaxs,prc,floatfix,amsmath]{revtex4-1}

\usepackage{graphicx} 
\usepackage{amsmath,amssymb}
\usepackage{esint}
\usepackage[utf8]{inputenc}
\usepackage{geometry}
\usepackage{xcolor}
\usepackage{float}
\usepackage{hyperref}
\usepackage{pgfgantt}
\usepackage{mathtools}
\usepackage[english]{babel}
\usepackage[utf8]{inputenc}
\newcounter{mycounter}
\fontsize{15}{13.2}
\begin{document}
\begin{center}

\textbf{Colliding and Fixed Target Mode in a Single Experiment--A Novel Approach to Study the Matter under New Extreme~Conditions}\\

\vspace*{0.5cm}
Oleksandr V. Vitiuk$^{1,}$ \footnote[1]{oleksandr.vitiuk@fys.uio.no},
Valery M. Pugatch $^{2}$, 
Kyrill A. Bugaev $^{3,4}$, 
Nazar S. Yakovenko $^{3}$, \\
Pavlo P. Panasiuk $^{3}$, 
Elizaveta S. Zherebtsova $^{5,6}$, 
Vasyl M. Dobishuk $^{2}$, 
Sergiy B. Chernyshenko$^{2}$,\\ 
Borys E. Grinyuk $^{4}$, 
Violetta Sagun $^{7,}$ \footnote[2]{violetta.sagun@uc.pt}, and
Oleksii Ivanytskyi $^{1}$ 

\vspace*{0.5cm}


$^1$\small{Institute of Theoretical Physics, University of Wroclaw, Max Born Pl. 9, 50-204 Wroc\l{}aw, Poland;}\\
$^2$\small{Institute for Nuclear Research, National Academy of Sciences of Ukraine, Prospekt Nauki av. 47, \linebreak 03680 Kyiv, Ukraine;} \\
$^3$\small{Department of Physics, Taras Shevchenko National University of Kyiv, 03022 Kyiv, Ukraine;}\\
$^4$\small{Bogolyubov Institute for Theoretical Physics, Metrologichna str. 14-B, 03143 Kyiv, Ukraine;}\\
$^5$\small{National Research Nuclear University (MEPhI), Kashirskoe Shosse 31, 115409 Moscow, Russia;}\\
$^6$\small{Institute for Nuclear Research, Russian Academy of Science, 108840 Moscow, Russia;}\\
$^7$\small{CFisUC, Department of Physics, University of Coimbra, Rua Larga, 3004-516, Coimbra, Portugal.}

\end{center}
\vspace*{0.5cm}

\begin{center}
{\bf Abstract} 
\end{center}
Here, we propose a novel approach to experimentally and theoretically study the properties of QCD matter under new extreme conditions, namely having an initial temperature over 300~MeV and baryonic charge density over three times the values of the normal nuclear density. According to contemporary theoretical knowledge, such conditions were not accessible during the early Universe evolution and are not accessible now in the known astrophysical phenomena. To achieve these new extreme conditions, we  proposed performing high-luminosity experiments at LHC or other colliders  by means of scattering the two colliding beams at the nuclei of a solid target that is  fixed at their interaction region. Under plausible assumptions, we estimate the reaction rate for the p+C+p and Pb+Pb+Pb reactions and discuss the energy deposition into the target and possible types of fixed targets for such reactions. To simulate the triple nuclear collisions, we employed the well-known  UrQMD 3.4 model for the beam center-of-mass collision energies $\sqrt {s_{NN}}$ = 2.76 TeV. As a result of our modeling, we found that, in the most central and simultaneous triple nuclear collisions, the initial baryonic charge density is approximately three times higher than the one achieved in the ordinary binary nuclear collisions at this energy.
%


\section{Introduction} \label{Intro}

During the last 30 years, progress in the  techniques and  methods in experimental nuclear and high energy physics has been impressive.
Probably the best illustration of this statement is the construction and performance of the Large Hadron Collider (LHC). The new technological  abilities allow us to hope that, soon, even more ambitious plans will come true. The International Linear Collider and  the Compact Linear Collider 
are the best examples of such ambitious plans. However, the financial and human  resources of the international community are limited and,  therefore, it will be difficult to simultaneously run the new projects together with the incomplete experimental programs that are running presently.  In other words, before starting to work on these ambitious plans, it is necessary to complete the major task of modern high-energy nuclear physics in the search for the quark deconfinement and (partial) chiral restoration phase transitions that are expected to occur in quantum chromodynamics (QCD). An additional aim is to either allocate their (tri)critical endpoint(s) or to disprove their existence. Below, we propose a new setup that could give the possibility to simultaneously probe high-temperature and high-density regimes and complete such a task.

After almost 40 years of experimental studies of the QCD phase structure in binary collisions (BC) of heavy atomic nuclei (for details see Refs.~\cite{PhysRevC.93.014902,PhysRevC.94.044912,EPJA2016,Sun:2017xrx,Moreau:2021clr} and references therein), several groups simultaneously realized  that,  besides the BC experiments, it is vitally necessary to have an independent, complementary  and reliable source of experimental information on the properties of QCD matter in order  to reach the goals of  these studi\-es~\cite{Adolfsson2020}.
Thus, astrophysical and gravitational waves probes of compact stars and their coalescence opened a possibility to study the equation of the state of QCD matter at very high baryonic densities~\cite{Ref1,Ref2}.  However, it provides a rather indirect  and very complicated way to obtain  information about the equation of the state of QCD matter. In addition, the phase transformations that occur in the BC experiments are strongly affected by the finite size of created systems, whereas, in binary neutron star mergers, finite size effects are not important~\cite{2019JPhG...46g3002O}. Therefore, the task to merge the information on the QCD matter equation of state obtained from these two sources might be highly non-trivial.

Understanding the possible difficulties, here, we suggest to use the triple nuclear collisions (TNC) as an independent and complementary  source  of information on the QCD matter equation of state. This is a novel approach that we suggest to refine further during the next years and employ it for constructing new targets, as well as for performing new kinds of experiments.
Despite the related extraordinary technological challenges that are also discussed, we would like to demonstrate that such experiments may open a window to probe the QCD matter under new extreme conditions, namely very high initial baryonic and electric charge densities, which cannot be achieved in the other experiments or any currently known natural phenomena. Hence, here, we discuss the p+C+p and Pb+Pb+Pb TNC to fully demonstrate the new prospects to study the properties of QCD matter.

{

To perform TNC simulations and calculate observables, we employed the microscopic transport model UrQMD-3.4~\cite{Urqmd1, Urqmd2}. Nowadays, the UrQMD model is one of the standard tools to simulate heavy-ion collisions as a purely hadronic model at low energies~\cite{UrqmdLowE} or as part of hydrodynamic-based hybrid approaches for the simulation of early and late non-equilibrium stages of the reaction~\cite{Urqmd3, UrqmdHybrid1, UrqmdHybrid2}. At the same time, this model can adequately reproduce LHC data in a purely hadronic regime~\cite{ALICEvsURQMD}. Therefore, we found the UrQMD model to be suitable for the TNC simulations, since it allows us to perform them without significant code changes.
}

The work is organized as follows. In Section~\ref{sec.2}, we estimate  the reaction rate of the
p+C+p and Pb+Pb+Pb triple nuclear collisions for the luminosity of proton and lead beams that will be available 
in the near future. In this section, we also discuss possible types of fixed targets for such experiments. 
Section~\ref{sec.3} is devoted to a discussion of the possible signatures of TNC.
In this section, we present the results of UrQMD 3.4 simulations 
of TNC and explain the transverse momentum redistribution effect, which can be important for 
the TNC detection in even-by-event analysis. In Section~\ref{sec.4}, the evolution of matter created in the central cell is studied, which is related to the study of the QCD phase diagram with the help of TNC. Our conclusions and a discussion of
perspectives are given  in Section~\ref{sec.5}.

%


\section{The Triple Nuclear Collisions Method}\label{sec.2}

Most high-energy physics experimental data nowadays are obtained in the controlled environment of hadron--nuclear colliders that are running in either fixed target or colliding regimes.  The main trends in the collider experiments are to 
increase luminosity (the  High Luminosity LHC, BELLE II)  or/and to increase the energy (International Linear Collider, FCC, etc.). In addition, there are successful fixed-target experimental programs such as NA49/NA61, HADES, BM@N, Fermilab~\cite{Fermilab} and others. On the other hand, there are interesting setups with the fixed target installed at the collider ring, and discussions are ongoing on the development directions of the fixed target regimes at LHC~\cite{AfterLHC1SM,AfterLHC2SM,AFTERsm} and RHIC~\cite{RHIC1sm}, inspired by the successful run of the fixed target setups at the storage ring HERA in the HERMES  experiment with a polarized gaseous target at the core of a positron beam~\cite{HERMESsm} and the HERA-B experiment with metal ribbons in a halo of the proton beam circulating with the energy $0.92$ TeV~\cite{HERAB1sm,HERAB2sm}. The fixed target experiments at RHIC and LHC allow for the high-energy nuclear physics community to explore new regions of the QCD phase diagram. We suggest to use this program and  to perform novel types of experiments on TNC, which can be  conducted in the colliding and fixed target modes, simultaneously. 
Note that the LHCb experiment has been successfully running the gaseous fixed target called System for Measuring the Overlap with Gas (SMOG) since 2013~\cite{LHCBdet,LHCBdetSMOG,SMOGNoble,LHCBdetSMOG2}.  The opportunity to increase the measured  experimental data by combining the fixed target mode with the colliding regime  was discussed  recently in Refs.~\cite{LHCBCol1,Pugatch1SM}.

Further below, we will propose the development of the TNC experiment conducted similarly by bombarding the solid-state fixed target with two colliding beams from the opposite sides. In addition, estimates for the TNC rates for various fixed targets will be presented below. 
The principal scheme of  dedicated studies with the LHCb detector  using a super-thin microstrip metal sensor~\cite{MMD1}, which can operate  either in a beam halo or in its core,  has been presented recently in~\cite{Pugatch2SM}.  
The new opportunities of the TNC method are also discussed in~\cite{Vitiuk2021TNC}.

Understanding the tremendous difficulties of the experiments on TNC, 
we will not discuss all technical aspects of such experiments in this work. Our primary goal is to emphasize that the potential of such experiments may initiate studies of entirely new regions of the QCD phase diagram that are not accessible by the other experimental techniques.  Nevertheless, 
as a starting point, below, we  estimate the rate of TNC for p+C+p and Pb+Pb+Pb collisions and discuss the basic properties of the solid fixed target necessary for such experiments on the TNC.

\subsection{Rate Estimates of the Triple Nuclear Collisions}

The concept of the TNC is just an extension of the usual high-energy nuclear collision of a pair of nuclei (A+A), with the third one being located in a fixed target at the interaction region of two colliding beams.

In Appendix \ref{apend}, we show the detailed calculations for the TNC rate for the p+C+p and Pb+Pb+Pb reactions for different arrangements of targets.

Thus, for the traditional arrangement of a target, the p+C+p and Pb+Pb+Pb TNC rate is equal to 
	\begin{equation} \label{Eqnew}
		\frac{d N_{p+C+p}}{d t} \simeq 4.8 \cdot 10^{-9} \frac{1}{s},~~~
		\frac{d N_{3Pb}}{d t} \simeq  2.4 \cdot 10^{-11} ~ \frac{1}{s},
	\end{equation}
considering the radius of nucleus consisting of A nucleons as $R_A \simeq 1.25 \cdot A^\frac{1}{3}$ fm.

At the same time, for the non-traditional arrangement of a target, which is discussed in detail in Section \ref{sub22}, we would expect a significant improvement in the reaction rate as
	\begin{equation} \label{EqpCp}
		\frac{d N_{p+C+p}}{d t} \simeq 2.0 \cdot 10^{-4} \frac{1}{s},~~~
		\frac{d N_{3Pb}}{d t} \simeq  3.4 \cdot 10^{-7} ~ \frac{1}{s}.
	\end{equation}

Moreover, an estimation of the energy deposition of two beams on the carbon target arranged non-traditionally is also presented in Appendix \ref{apend}.

\subsection{Discussion of Different Types of Fixed Targets}
\label{sub22}

In this subsection, we discuss and evaluate a possible  experimental model for TNC.
Up until now, the only experiment available to gather data in both collider and fixed-target modes is LHCb. Its fixed target system SMOG~\cite{LHCBdetSMOG,SMOGNoble} was originally developed for a precise luminosity calibration of colliding proton beams and allows one to inject a low flow rate of noble gas into the vacuum vessel of the vertex detector. The thickness of the SMOG gaseous target of approximately $\tau \simeq 10^{12}$ atoms$\cdot$cm$^{-2}$~\cite{SMOG2} was suitable to run the fixed target mode in parallel to the colliding mode. In addition, its upgraded version, with a capability to increase a target density by an order of magnitude, i.e., up to $10^{13}$ atoms$\cdot$cm$^{-2}$, is installed for the RUN3 on LHC~\cite{SMOG2}.

Another option is the use of the solid fixed target (SFT). Utilizing SFT for the studies of the TNC allows us to introduce a variety of nuclei for the experiments and, also, to carefully select a wide range of nuclei parameters, such as shape deformation, neutron excess, spin and parity, for precise studies of strongly interacting matter created during the nuclei collisions.

Technically, the SFT mode can also provide a number of  such advantages as the extremely precise positioning of primary vertices, suitable tuning of the interaction rate and a simple change in the target. 
However, the highest  TNC rates given by  Equation~(\ref{EqpCp}) for the non-traditional arrangement of the target 
may create  other problems, even in the case where the large energy deposition by two high luminosity beams on the target is solved by fast rotation of~the target.
 
The whole point is that the huge fluxes of radiation may cause quenching of superconductive magnets, as well as possibly 
producing an
irreversible radiation damage of the detector elements. Hence, one has to find  a solution in tuning the experimental features that would be suitable for the TNC observation yet tolerable for other systems.
One apparent way is to enhance the characteristics of beams: a better focusing of the beams to reach the  submicron  value of the beam core radius could reduce the radiation loads and increase the luminosity;
a reduction or bunch length  and synchronization in the picosecond or even femtosecond domain. Note that the X-ray free-electron laser facility  (XFEL) at  DESY  recently demonstrated a beam pulse of  a few fs~\cite{XFEL}. It is important to note that LHC already obtained an increase in luminosi\-ty due to the decrease in the beam core radius.
There are also considerations~\cite{XFEL}  to move the studies at the XFEL to the attosecond domain, 
i.e., to a beam pulse of $10^{-18}$ s.

Now, assume that the improvement in beam characteristics and increasing  luminosity can increase  the TNC rate by two orders of magnitude. Then, one can consider the following options for SFT.

{\bf Option SFT1.} A  {\it super-thin SFT made of the gra\-phene layer} of thickness $l_g \ll 3.32 \, \mu$m. In this case, the TNC rate of Equation~(\ref{Eqnew}) has to be multiplied by the factor 
$100 \cdot \frac{l_g}{2 R} \frac{1.43}{2.2}$, and we obtain $\frac{d N_{p+C+p}}{d t}_{graphene} \simeq  \frac{l_g}{2 R} \cdot 10^{-5} \, {\rm s}^{-1}$ for the p+C+p collisions. We have taken into account the different  target thicknesses and the fact that graphene has a density of $1.43$ g$\cdot$cm$^{-3}$, which is essentially  lower than carbon. Evidently, a too thin target may significantly reduce the rate of the TNC. Apparently, such a target can be used just for testing  the new  techno\-lo\-gy for TNC measurements, since the resulting  reaction rate is still low.

{\bf Option SFT2.} A {\it rotating and restorable SFT}. In our estimates of the reaction rate of Equation~(\ref{Eqnew}) for the traditional assembling of the target, we assumed the geometrical thickness of the target to be $l_g = 3.32 \, \mu$m.  However, our analysis of the energy deposition to the target shows that the geometrical thickness  target $l_g $ can be increased by a factor of 30 or even 100, i.e., one can 
employ 
 $l_g \in 100-332 \, \mu$m if it rotates  with the linear speed of 30--100 m/s.  In this case, one can expect an additional increase in
 the reaction rate by a factor of 30--100, which, in total, gives us a sizable enhancement of the reaction rate (\ref{Eqnew}) by 3000--10,000 times.
The latter estimate seems to be  feasible for the TNC detection, not only for the p+C+p reactions, but also for 
the Pb+Pb+Pb ones. However, for the  TNC experiments  with heavy nuclei,  it is perhaps more realistic to consider a stronger target 
made of bismuth (Bi) or tungsten (W).

In addition, it is clear that, for the options SFT1 and SFT2, one has to provide a sufficiently  long  time of the target operation. 
We believe that the restorable targets can be designed using  the  micro-electromechanical systems (MEMS) technology.
This is an advanced technology in the fixed target  engineering that can provide the micrometer  movements  of the  target near the beam with a nanometer precision~\cite{Barschel:2020drr}.

{\bf Option SFT3.} A {\it jet micro-powder (pellet)  target} successfully operated at the electron storage ring~\cite{SFT1a} with its target thickness for the Ni micro-powder particles, with a diameter of approximately 1 $\mu$m being approximately $ \tau_{Ni} \simeq 10^{16}$ atoms$\cdot$ cm$^{-2}$~\cite{SFT1b}. Taking into account recent technological advances, one can hope that such targets will be revisited with a significantly increased density of jet micro-powder for the Pb, Wi or  W particles. Moreover, it might be possible for the modern jet micro-powder targets to provide a sufficiently higher speed of particle flow to essentially reduce the energy deposition effects discussed above.

{\bf Option 4.}  The cardinal solution to this problem could be a construction of 
a (vertically oriented)  {\it storage ring with a high-intensity  ion beam}  that is  focused toward the interaction region  of two collider beams with a submicron size and a similar positioning accuracy. Even  the expected density of  an ion beam  of  $10^{21}$ ions/cm$^3$  can  be used for the TNC experiments in the nearest future,  while one can hope that, in a few years,  this density of ions in the beam can be 
increased further.

Although there are many unclear issues and unfixed problems of the TNC, we hope that, from the discussion above, one can straightforwardly conclude that the SFT experiments used to generate the TNC are not the theo\-re\-tical dreams of today, but a technological reality of tomorrow.

\section{Possible Signatures of TNC}\label{sec.3}

\begin{figure}[H]
\hspace*{-30.2mm}
\includegraphics[width=0.5\columnwidth]{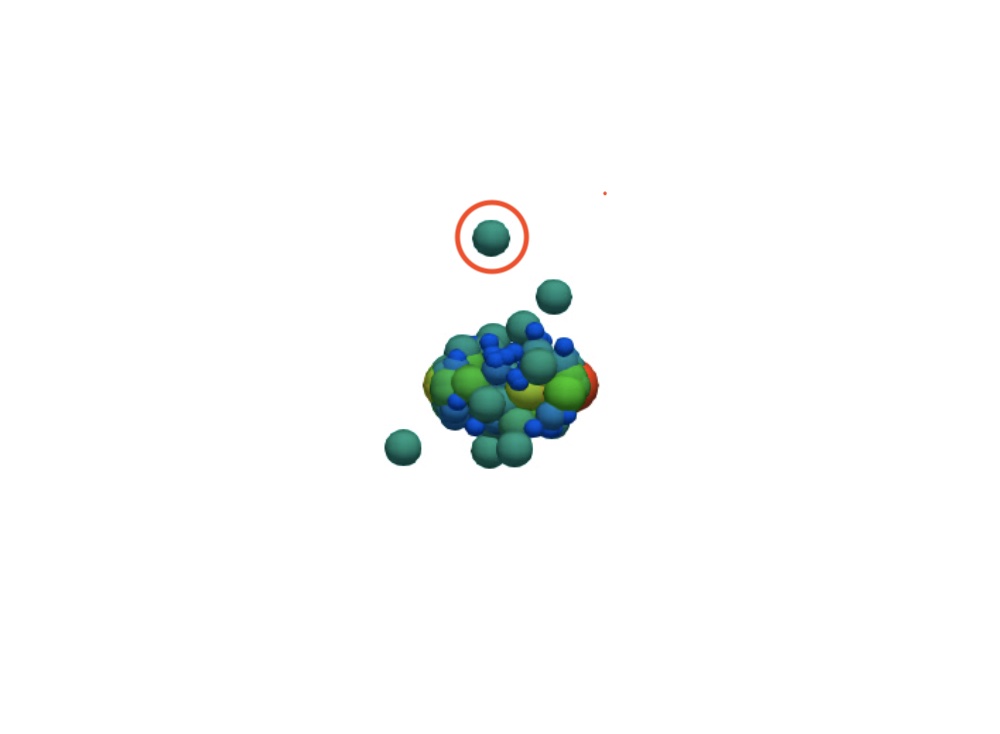} \hspace*{-19.2mm}
\includegraphics[width=0.5\columnwidth]{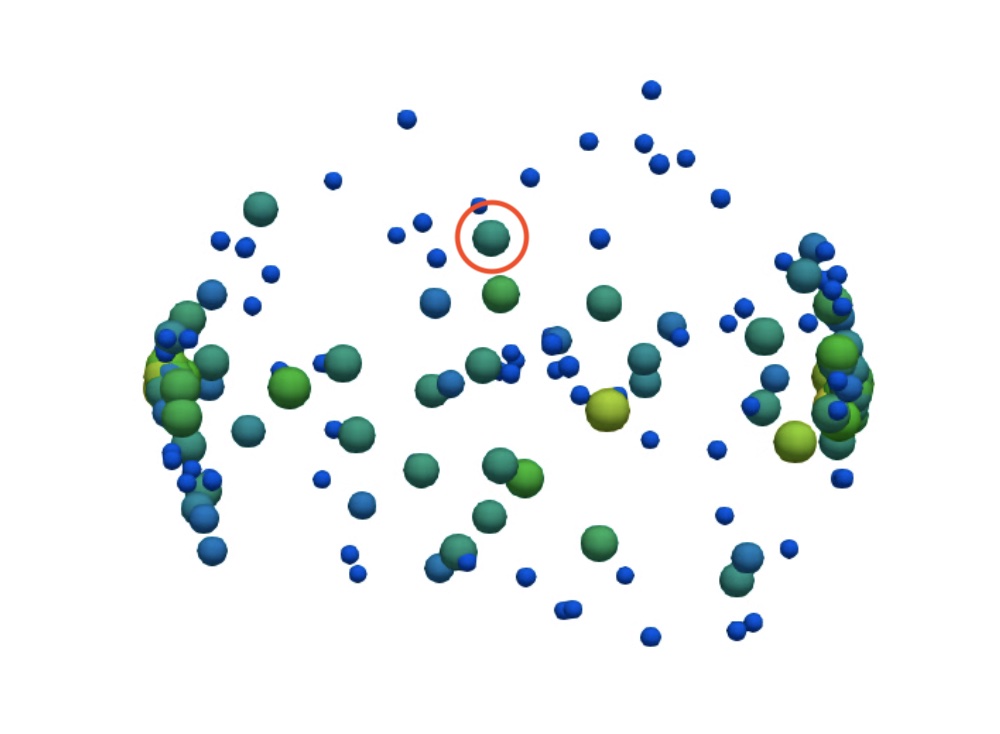}
\caption{Schematic picture of the p+C+p TNC at the center-of-mass collision energy of protons $\sqrt{s} = 2.76$ TeV. It visualizes the results of the UrQMD 3.4 simulations for $t = 5$ fm  (\textbf{left panel}) and for  $t = 8.4$ fm  (\textbf{right panel}) after first collision of protons. The volume $V_h$ of particles is proportional to their mass  $m_h$ for a better perception. The encircled nucleon shown in both panels does not  leave its position during the reaction.}
\label{Fig1}
\end{figure}
\begin{figure}[H]
\includegraphics[width=0.51\columnwidth]{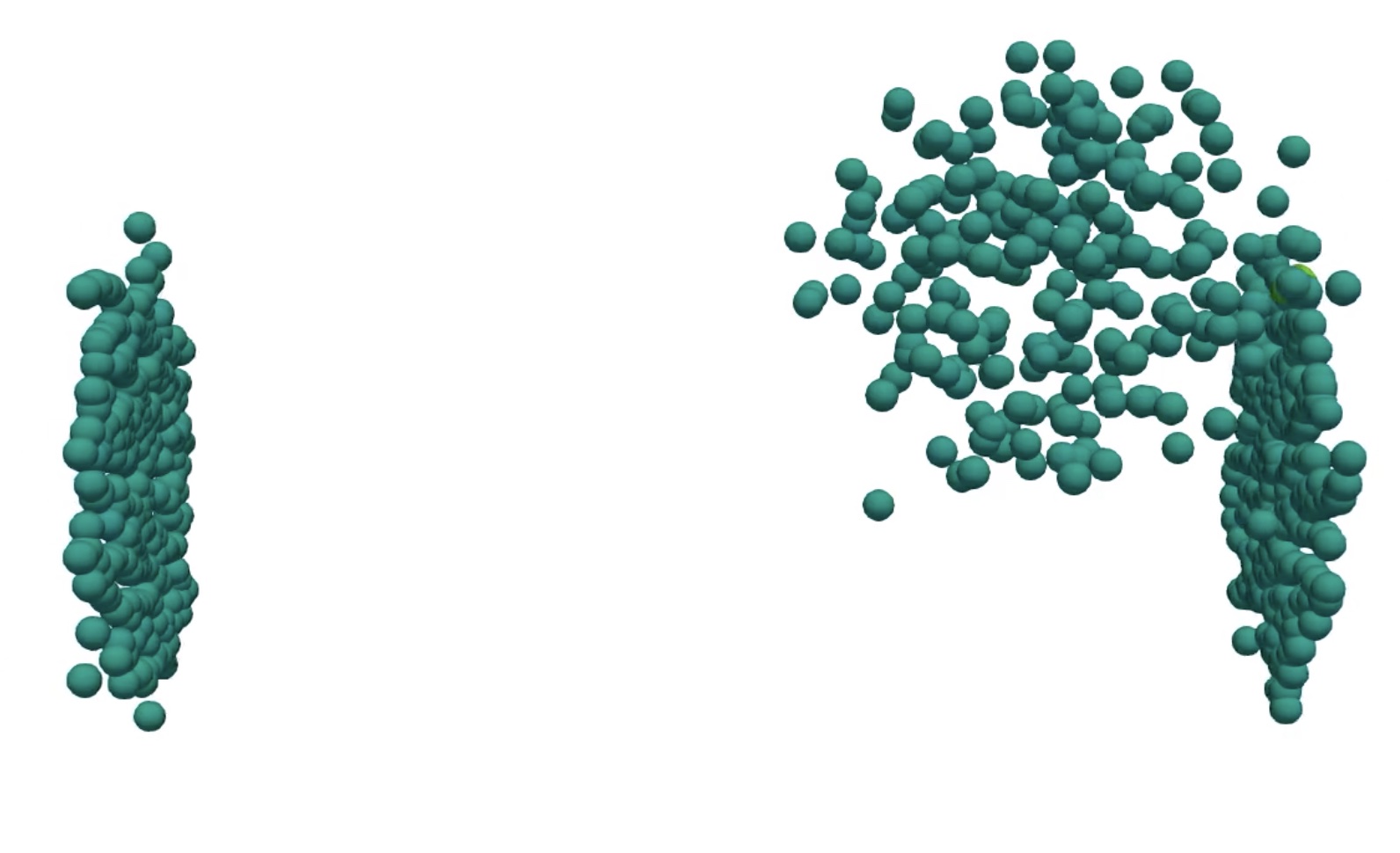} 
\hspace*{-3.2mm}
\includegraphics[width=0.51\columnwidth]{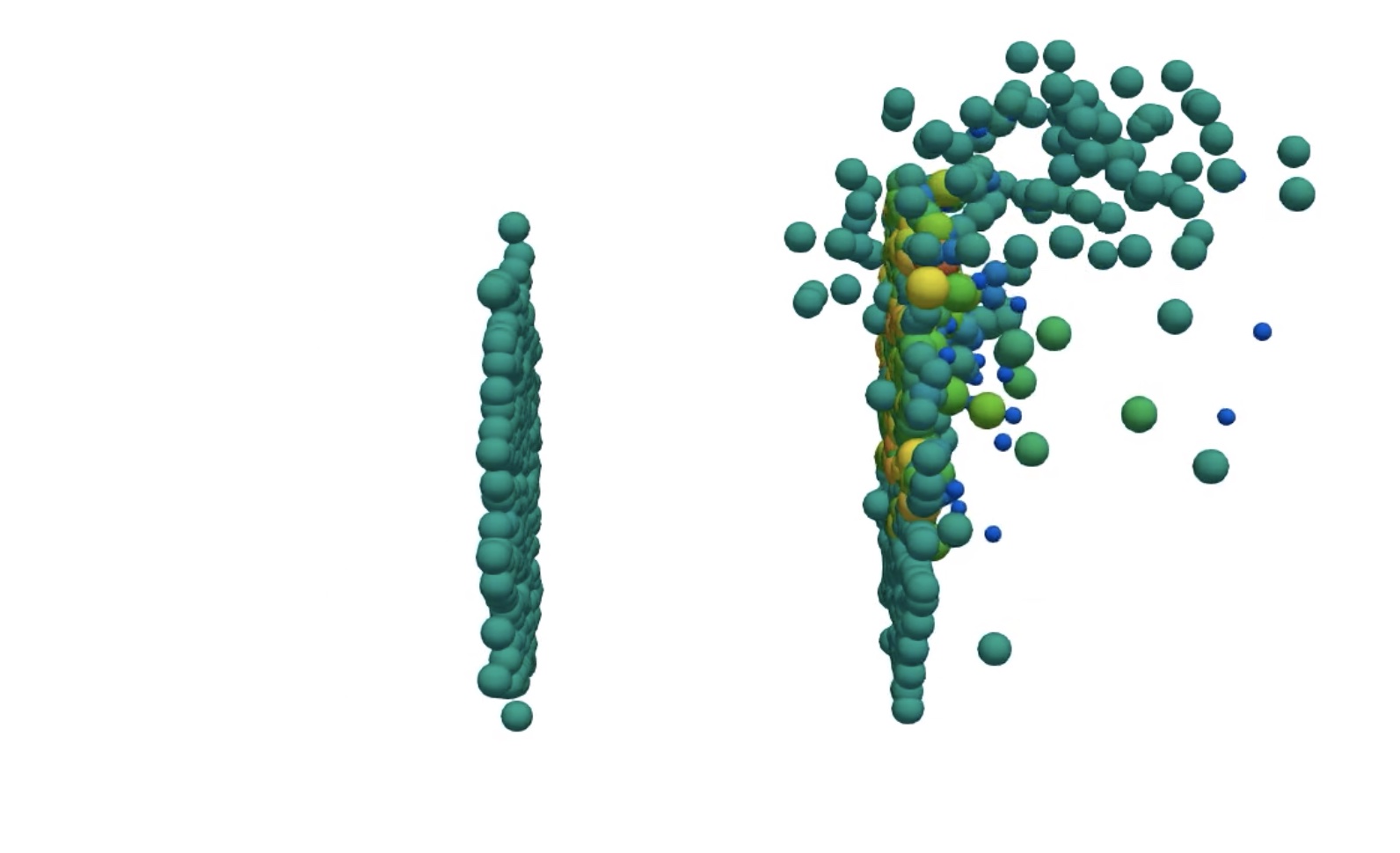}
\caption{Schematic picture of the non-central Pb+Pb+Pb TNC with time delay $t_{del} =10$ fm at the center-of-mass collision energy of beam ions $\sqrt{s} = 2.76$ TeV. It visualizes the results of the UrQMD 3.4 simulations:  the beginning  of collision of target nucleus with the right ion (\textbf{left panel}) and  the moment when the right ion almost passed through the target nucleus (\textbf{right panel}). The volume $V_h$ of particles is proportional to their mass  $m_h$ for a better perception.}
\label{Fig2}
\end{figure}

To demonstrate new opportunities that the TNC me\-thod opens to investigate the QCD phase diagram, we employed the Ultra-relativistic Quantum Molecular Dynamics (UrQMD) transport model~\cite{Urqmd1,Urqmd2} (version 3.4) to simulate simultaneous central p+C+p and Pb+Pb+Pb reactions. Some typical examples of the TNC simulations within the UrQMD model are shown in  Figures~\ref{Fig1} and \ref{Fig2}. To adapt the UrQMD model for the TNC simulations setup, we used two colli\-ding nuclei (or protons) with a zero impact parameter, e.g., moving along z-axis. The initial distance between these nuclei (protons) was increased to be sufficient for fitting the third nucleus between them. After that, the third nucleus, whose center coincides with the origin of the coor\-dinate system, was added between two colliding beam nuclei at the initialization stage. This means that, here, we studied a particular case of central simultaneous TNC. The snapshot of more realistic events is shown in Figure \ref{Fig2}.
For many years, the UrQMD~\cite{Urqmd1,Urqmd2} has served as a reliable tool to predict and describe the A+A collision experiments~\cite{Urqmd3,Urqmd4}. {Although UrQMD is successfully used even for the top RHIC collision energy, interpretation of the results of BC simulations at the LHC energy $\sqrt{s_{NN}} = 2.76$ TeV with the UrQMD should be performed with care, as a purely hadronic transport model may be not sufficient to describe the dynamics of the hot and dense stage of heavy-ion reactions at high energies~\cite{Urqmd3}.} Therefore, first, we compared the results of UrQMD simulations obtained for the p+p and Pb+Pb collisions at $\sqrt{s_{NN}} = 2.76$ TeV with the experimental data. 
\vspace{-14pt}

\begin{figure}[H]
\includegraphics[width=0.5\columnwidth]{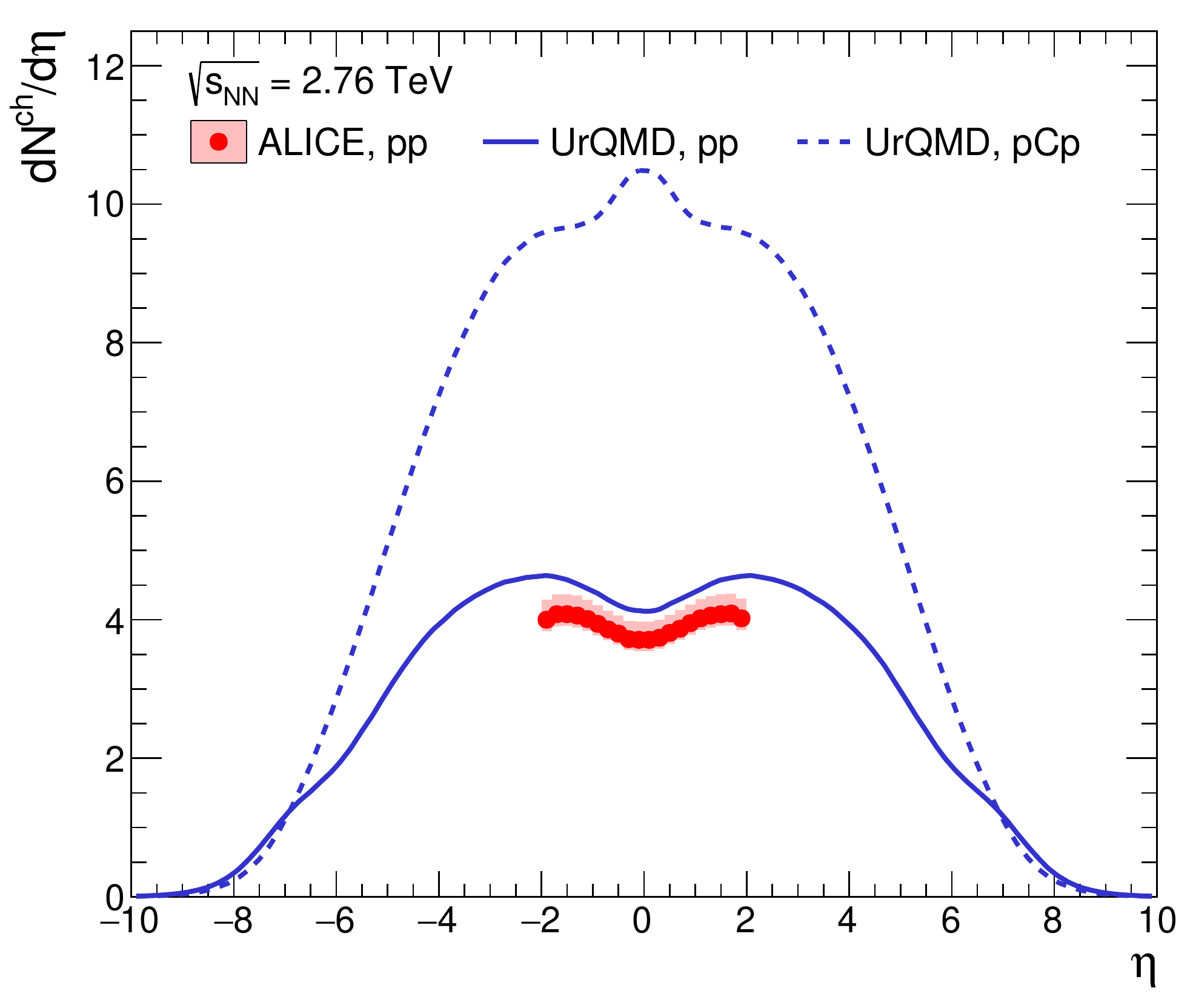}
\includegraphics[width=0.5\columnwidth]{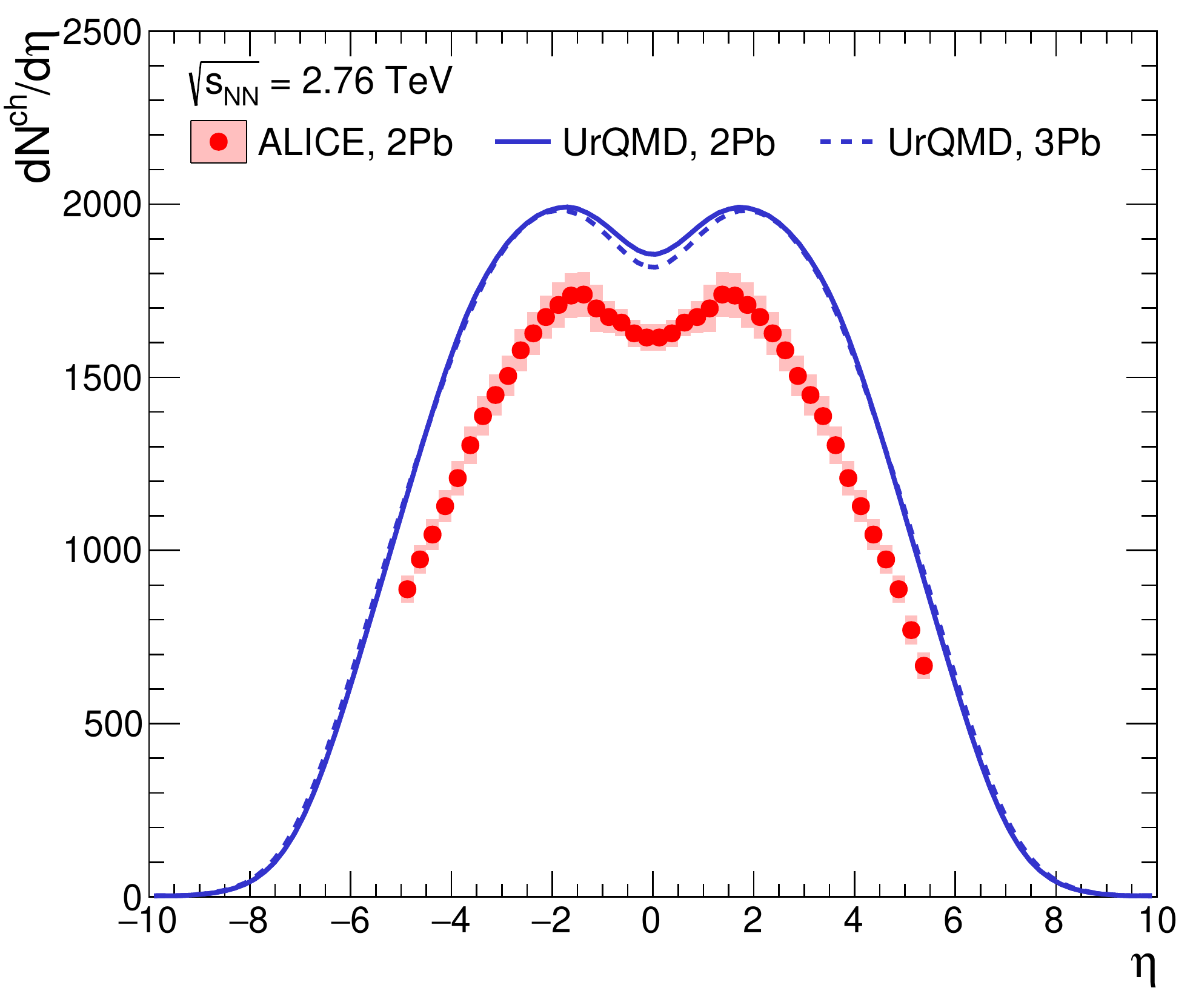}
\caption{Comparison of the pseudorapidity distributions of charged particles $\frac{d N_{ch}}{d \eta}$ found in the experiments with the ones obtained by the UrQMD simulations. ({\bf Left panel:}) ALICE data on inelastic p+p collisions of Ref.~\cite{ALICE_pp_eta} measured at minimum bias at $\sqrt{s_{NN}} = 2.76$ TeV (symbols) vs. the UrQMD results (solid curve). For comparison, the dashed curve shows the same distribution for the p+C+p TNC. ({\bf Right panel:}) ALICE data of Ref.~\cite{ALICEvsURQMD} measured in 0--5\% most central Pb+Pb collisions at $\sqrt{s_{NN}} = 2.76$ TeV (symbols) vs. the UrQMD results (solid curve). The dashed curve shows the same distribution for the Pb+Pb+Pb TNC.}
	\label{KAB_Fig1.3}
\end{figure}

As one can see from Figure~\ref{KAB_Fig1.3}, for both reactions at the pseudorapidity values $|\eta| \le 1$, the deviations in the UrQMD results for the $\frac{d N_{ch}}{d \eta}$ distribution of charged particles are, respectively, approximately 10\% for the p+p collisions and approximately 17\% for the Pb+Pb BC, although, in both cases, we used a zero value of the impact parame\-ter $b=0$, while ALICE data corresponded to 0--5\% centrality class.
Our results obtained with UrQMD for the Pb+Pb BC are similar to the results of Ref.~\cite{ALICEvsURQMD} (see Figure 6 in there).

\begin{figure}[H]
	%
	\includegraphics[width=0.5\columnwidth]{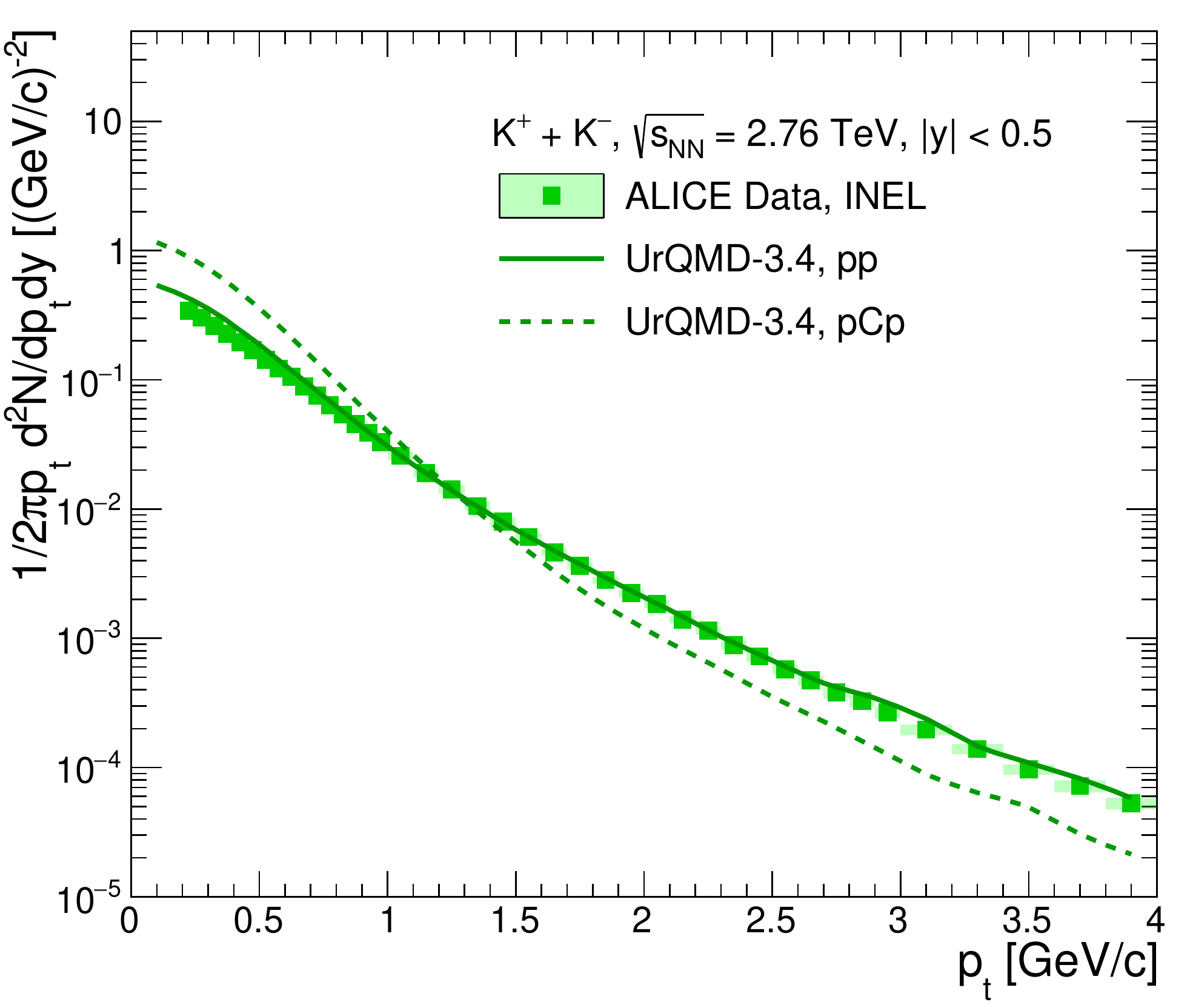}
    \includegraphics[width=0.5\columnwidth]{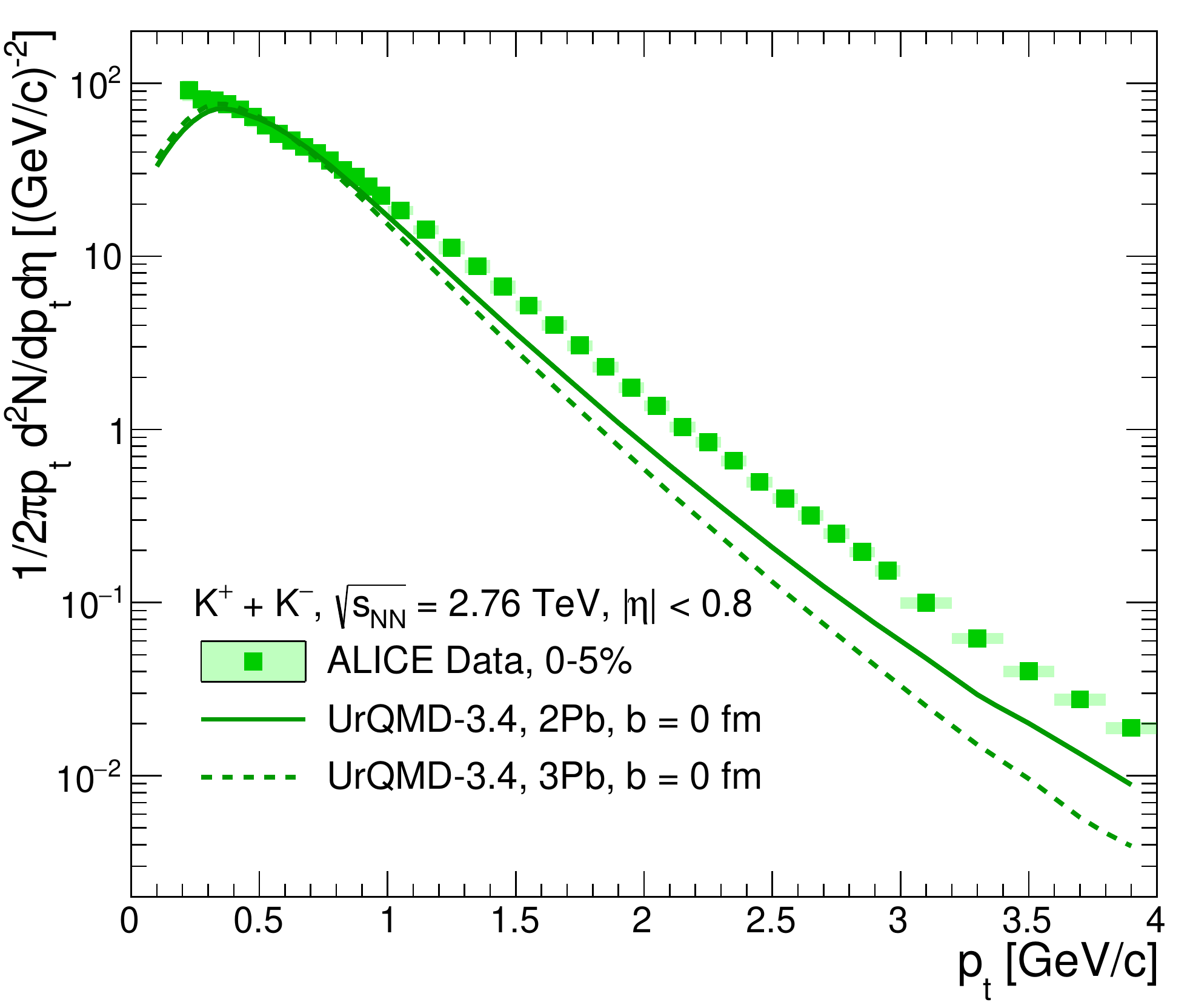}	
	\vspace*{-2mm}
	\caption{Comparison of the $p_T$ spectra of kaons found in the experiments with the ones obtained by the UrQMD simulations. The solid curve shows the results of BC, whereas the dashed curve shows the results for TNC.
		({\bf Left panel:}) ALICE data on inelastic p+p collisions of Ref.~\cite{ALICE_pp_pT,ALICE2013,ALICE2016} measured at minimum bias (symbols) vs. the UrQMD results for the p+p BC and for the p+C+p TNC.
		({\bf Right panel:}) ALICE data of Ref.~\cite{ALICEvsURQMD} measured in 0--5\% most central Pb+Pb collisions vs. the UrQMD results for the Pb+Pb BC and for the Pb+Pb+Pb TNC.}
	\label{KAB_Fig2.3}
\end{figure}

In addition, for illustrative purposes, in Figure~\ref{KAB_Fig2.3}, we demonstrate the $K^++K^-$ transversal spectra for both reactions for the most typical ones. In general, the agreement with data is satisfactory, although, for the Pb+Pb BC, the deviation at $p_T > 1.2$ GeV is quite sizable, but this part of the $p_T$ spectrum gives only a small part ($<$ 3\%) of the total particle yield. 
Nevertheless, the comparison shown in Figures~\ref{KAB_Fig1.3} and \ref{KAB_Fig2.3} justifies our usage of UrQMD to study, with its help, the initial baryonic charge density and the bulk features of hadron production in BC and in TNC at LHC collision energies.
At the same time, the phase transformations occurring in such collisions should be studied with the more speciali\-zed  models and, hence, we do not discuss them here.

In order to reduce the effects of the discrepancy bet\-ween the UrQMD results and the experimental data, here, we analyzed the ratios of quantities found in the TNC to the ones found in the BC, i.e., the 3-to-2  modi\-fication  factor.
We hope that the deviation in calculated quantities will be greatly reduced for such ratios, since the source of deviations is the same in either type of reactions.
Another reason to demonstrate the 3-to-2 modification factor is that, in log-scale, the difference in the momentum spectra obtained for the TNC and for BC is not seen clearly, while the 3-to-2  modification factor shows this difference very clearly.

Here, we define the 3-to-2  modification factor of hadronic transverse momentum spectra~as
\begin{eqnarray}
  R_{32} = \left(\dfrac{1}{2\pi N_{\text{events}}} \dfrac{d^2N}{p_tdp_tdy}\right)_{\text{TNC}}\Bigg/\left(\dfrac{1}{2\pi N_\text{events}} \dfrac{d^2N}{p_tdp_tdy}\right)_{\text{BC}}.
    \label{EqR32}
\end{eqnarray}

\begin{figure}[H]
	\centerline{
	\includegraphics[width=0.5\columnwidth]{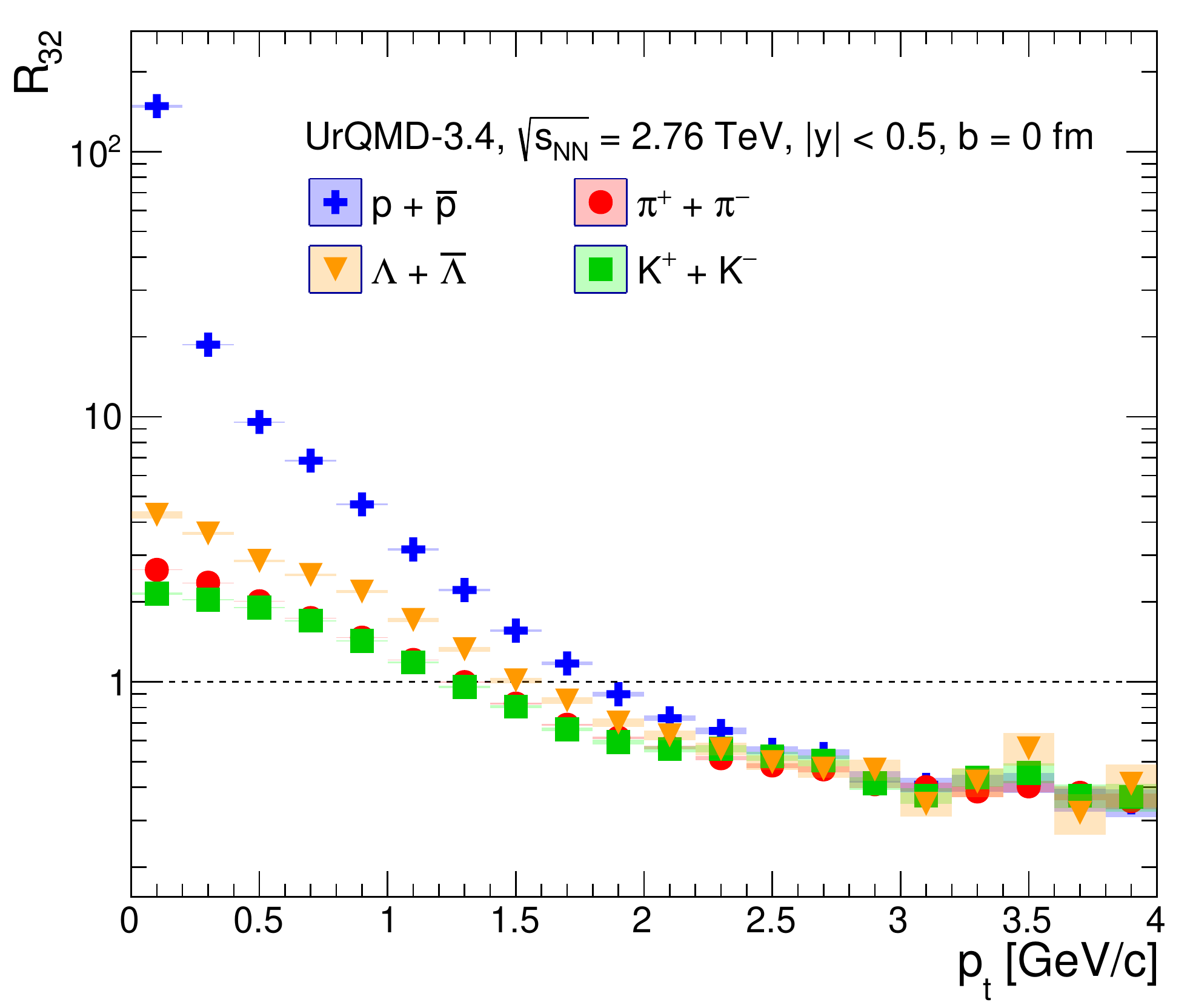}
	\includegraphics[width=0.51\columnwidth]{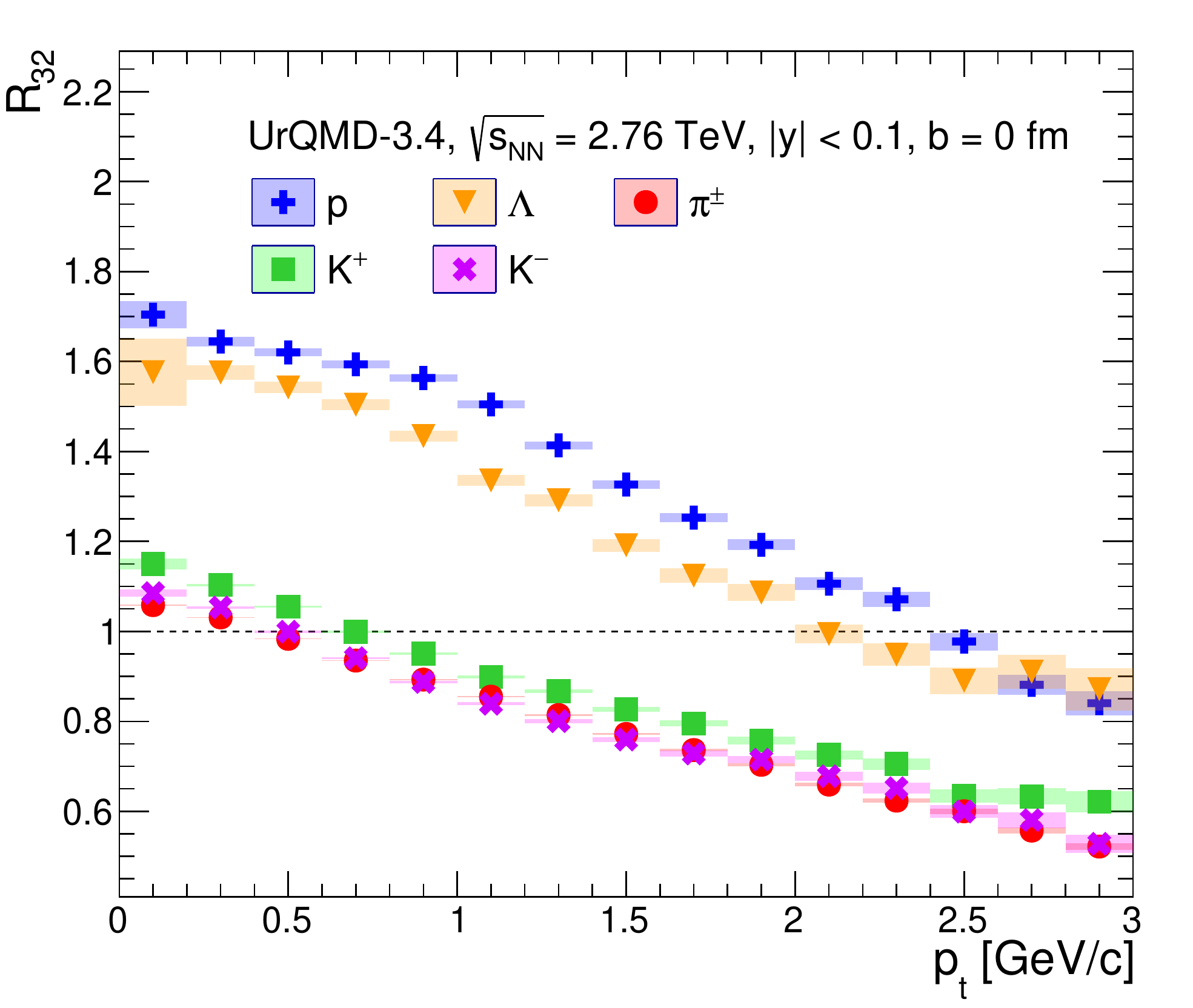}
	}
	\vspace*{-2mm}
	\caption{Ratio of transversal momentum spectra of TNC to the one of A+A collisions (3-to-2 nuclei modification factor) of hadrons obtained for the same collision energy $\sqrt{s_{NN}} = 2.76$ TeV as a function of particle transverse momentum with the ones obtained by the UrQMD simulations.
		({\bf Left panel:}) Results for the p+C+p TNC are found at $\left|y\right| < 0.5$.
		({\bf Right panel:}) Results for the Pb+Pb+Pb TNC are found at $\left|y\right| < 0.1$.}
	\label{KAB_Fig3.3}
\end{figure}

The advantages of using $R_{32}$ are apparently seen if one compares Figures~\ref{KAB_Fig2.3} and \ref{KAB_Fig3.3}. From the left panel of Figure~\ref{KAB_Fig3.3}, one can immediately see two strong effects: first, there is a strong 
(especially for protons) enhancement of particle $p_T$ spectra in the TNC at low values of transversal momentum $p_T < 0.5$ GeV/c; second, for $p_T > 2$ GeV/c, there exists a suppression of particle $p_T$ spectra found in the TNC. 
An apparent explanation of these effects is related to the fact that, in the p+C+p TNC, the nucleons of the carbon target nucleus exist in the central zone of the reaction and a few of its remnants (one or two nucleons) remain there even after the reaction (compare the left and right panels of Figure~\ref{Fig1}). Due to the presence of surrounding nucleons, the products of the primary p+p or p+n collisions moving in the transversal direction experience re-scattering on the nucleons of the target, lose their momenta and produce an additional number of hadrons compared to the case of ordinary BC. As a result, the probability of finding slow-moving nucleons at the midrapidity strongly increases, while, due to one or two re-scatterings, the products of primary collisions with large $p_T$ lose part of their momentum and populate the states with low $p_T$ values.
Therefore, these features can be used to identify the p+C+p TNC, measuring the following ratio:
\begin{eqnarray}\label{Eq14vn}
	R ({\rm low~ p_T~ to~ high~ p_T}) \equiv \frac{d^2 N }{p_T d p_T d y} \biggl|_{p_T < 1 \rm GeV/c} \Biggl/ \frac{d^2 N }{p_T d p_T d y} \biggl|_{p_T > 2.5 \rm GeV/c} 
\end{eqnarray}
for protons and $\Lambda$-hyperons (and, maybe, kaons) in each event and comparing it with the averaged one.

Qualitatively the same picture can be observed in the Pb+Pb+Pb TNC, as is seen in the right panel of Figure~\ref{KAB_Fig3.3}.
However, due to many re-scatterings, the effect of a low $p_T$ enhancement weakens essentially for pions and kaons, whereas it remains sizeable only for protons and $\Lambda$-hyperons. On the other hand, a sizeable suppression of pions and kaons with the transversal momenta above 1.5 GeV/c is clearly seen in Figure~\ref{KAB_Fig3.3}, whereas, for protons and $\Lambda$-hyperons, the high $p_T$ suppression appears at $p_T > 2.5$ GeV/c.
Therefore, we suggest to call it the transverse momentum redistribution effect.
We would like to note that, in the case of pions and kaons, this redistribution enhances the low momentum modes and can potentially lead to overpopulation of the zero momentum one, i.e., to the formation of the corresponding Bose--Einstein condensates (BEC), which recently were discussed in the context of LHC energies---for a recent example, see~\cite{BEC2021}. If this is the case, then the p+C+p and Pb+Pb+Pb TNC can be the best candidates to observe the {BEC} of pions or/and kaons. At the same time, we stress that the absence of the effects of quantum statistics of particles in the UrQMD 3.4 spoils the understanding of the physical reasons of the mentioned low $p_T$ enhancement and requires further study.

Similarly to the case of p+C+p TNC, the transverse momentum redistribution effect can be used to identify the Pb+Pb+Pb TNC using the event-by-event analysis. In addition, this effect leads to the smaller charged particle number at mid-rapidity as is shown in Figure~\ref{KAB_Fig3.3}. Due to the fact that the average transverse momentum of particles produced in TNC is lower than in the BC, they are redistributed from the mid-rapidity region to the tails of the distribution. At the same time, the total number of produced particles is approximately the same in the BC and TNC setups.

\begin{figure}[H]
	\centering
\includegraphics[width=0.5\columnwidth]{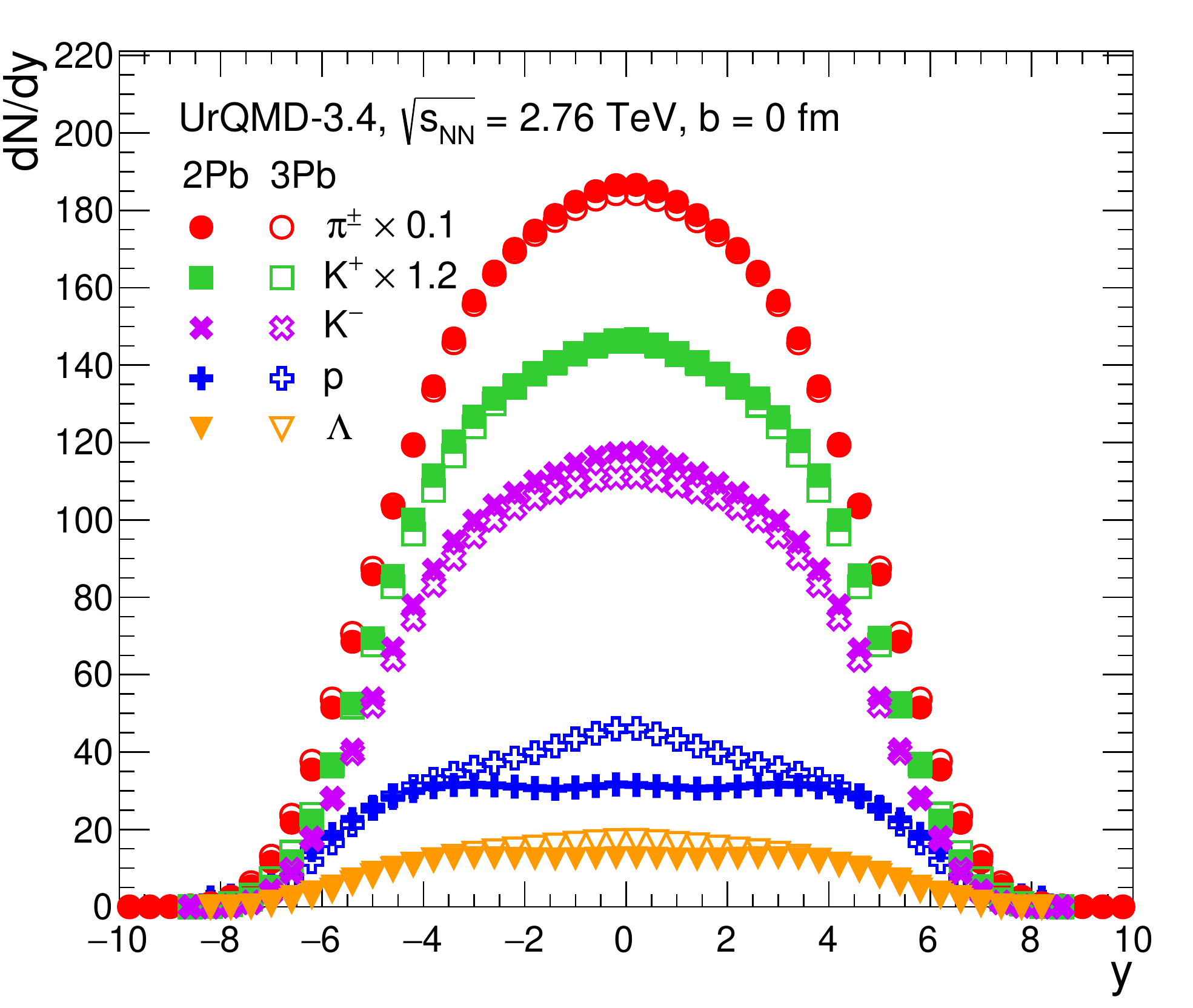}
	\caption{Comparison of the hadronic yields found for the central Pb+Pb BC (filled symbols) and for the central-simultaneous Pb+Pb+Pb TNC (open symbols) as the function of particle rapidity $y$.}
	\label{KAB_Fig4.3}
\end{figure}

The other possible way to identify the Pb+Pb+Pb TNC is related to approximately a 50\% enhancement of the proton yield at midrapidity compared to the BC, as one can see from Figure~\ref{KAB_Fig4.3}, and a slightly weaker enhancement of the $\Lambda$-hyperon yield.
In order to look into more details on the evolution of a system created in the Pb+Pb+Pb TNC, below, we study the time dependence of the thermodynamic quantities of the central cell.

However, all of these approaches based on the event-by-event basis are difficult to realize because of the very low ratio of the TNC signal to the BC background. Therefore, we also propose an approach that is based on the particle rapidity distributions and primary vertex location.
First of all, rapidity distributions of particles created in the fixed target mode collisions will have a peak near $\pm y_{beam}$. At the same time, rapidity distributions of particles produced in the BC and TNC will have a maximum at zero rapidity, which allows us to separate them from fixed target events easily. To separate collider mode BC from TNC, let us consider that one uses a fixed target in the form of a sequence of thin films. With such a setup, one can see if the interaction vertex is inside of the film or not. If the films are thin enough, then binary collisions will happen mostly in the space between films, and only TNC will be inside the film (fixed target events are already subtracted using rapidity distribution), or at least one can reduce a background from collider BC by removing all interactions that are outside of films.

\section{Evolution of Matter in Central Cell in Pb+Pb+Pb Collisions}\label{sec.4}

A rather intriguing question arises: what regions of the QCD phase diagram can one probe with the TNC? To answer it, we studied the time evolution of the net baryonic charge density in the central cell of the volume 27 fm$^3$. In fact, we considered the range of cell volumes from $1$ fm$^3$ to $27$ fm$^3$ and chose this value since the obtained results become the volume-independent one if the volume of the central cell is above 6 fm$^3$. Moreover, to study the collision energy dependence, we analyzed the Pb+Pb+Pb TNC for two energies of the collision: $\sqrt{s_{NN}} = 2.76$ TeV (LHC) and $\sqrt{s_{NN}} = 200$ GeV (top RHIC energy).

\begin{figure}[H]
\centering
{\includegraphics[width=0.5\columnwidth]{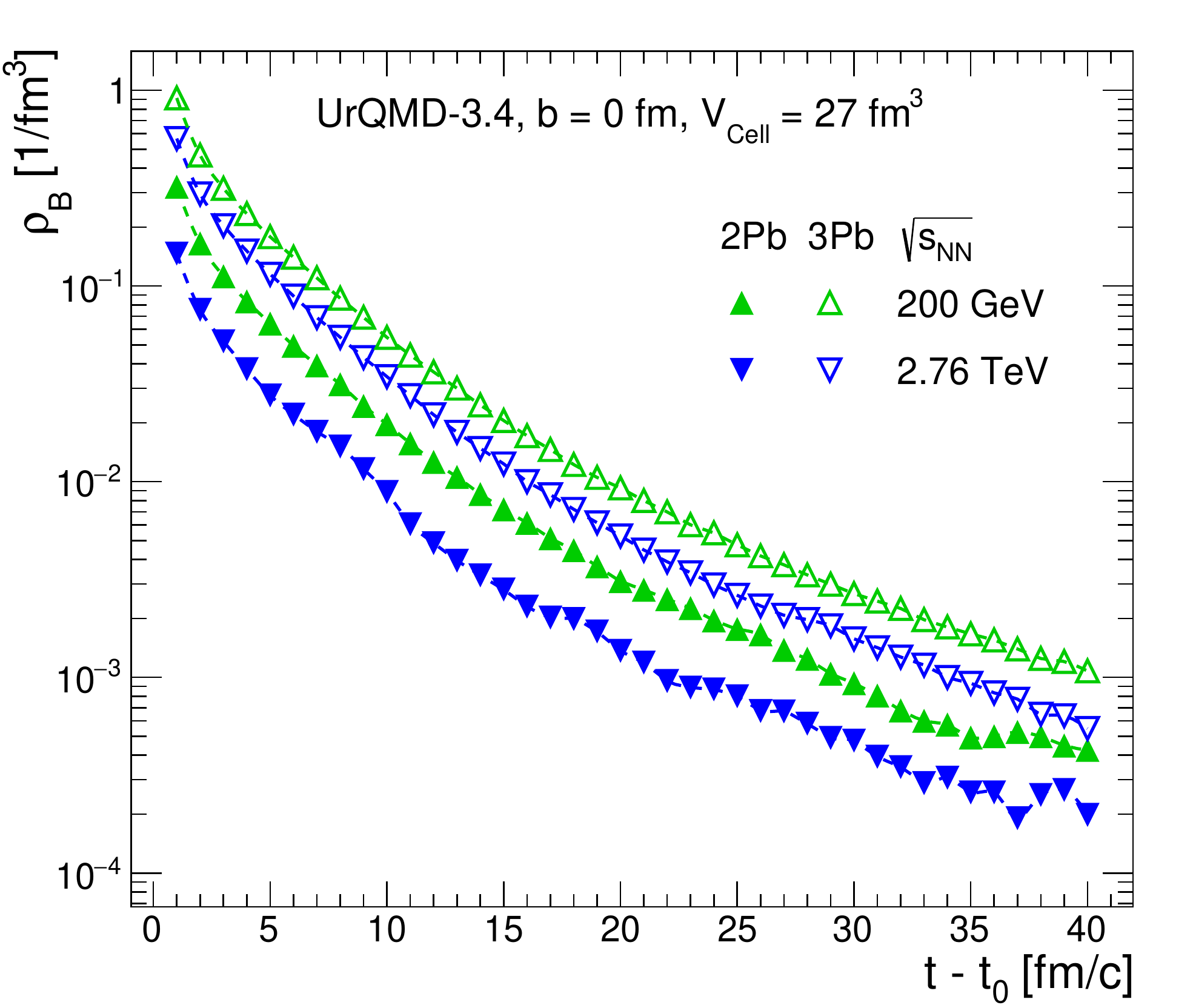}}
	\caption{The time evolution of the central cell baryonic charge density of the volume $27$ fm$^3$ during the process of ordinary BC collisions (filled symbols) and the TNC (empty symbols) for $\sqrt{s_{NN}} = 200$~GeV (triangles pointing upwards) and for $\sqrt{s_{NN}} = 2.76$~TeV (triangles pointing downwards). The time $t_0$ is the moment at which the remnants of projectile nuclei have passed through the central cell.
	}
	\label{KAB_Fig1.4}
\end{figure}

As one can see from Figure~\ref{KAB_Fig1.4}, in comparison to the usual BC, the TNC provides a very large increase in the initial net baryonic charge density in the TNC: from $0.321$ fm$^{-3}$ to $0.913$ fm$^{-3}$ for $\sqrt{s_{NN}} = 200$ GeV and from $0.146$ fm$^{-3}$ to $0.567$ fm$^{-3}$ for $\sqrt{s_{NN}} = 2.76$ TeV. These results show us that, in the TNC, one can expect a formation of three times denser initial states of net baryonic charge for RHIC top energy and four times denser ones for LHC energy than in the BC. 
Such an increase in the baryon charge density is mostly caused by involving more wounded nucleons in the TNC compared to the BC. At the same time, the volumes containing these nucleons at the initial moment of TNC and BC are close since they are roughly equal to the volumes of the overlap of the colliding nuclei. Since the fixed target nucleus is in the rest, its contribution to the energy deposit is negligible in comparison with the one of the two beam nuclei. Thus, the energy densities that can be achieved in the TNC and BC are very close due to the similar volume of the reaction fireball and energy deposit.

It should be mentioned here that the time $t_0 \simeq 10$ fm/c for the Pb+Pb+Pb TNC and $t_0 \simeq 1$ fm/c for Pb+Pb collisions correspond to the moment at which the remnants of target nuclei have passed through the central cell (see Figure~\ref{KAB_Fig1.4}).

In order to demonstrate  the increase in baryonic chemical potential that can be achieved in the Pb+Pb+Pb TNC, we used the MIT bag model equation of state~\cite{BagModelSM} for three quark flavors and three colors. For the massless quarks, the MIT bag model pressure as the function of system temperature $T$ and baryonic chemical potential $\mu_B$ can be cast as
\begin{eqnarray}\label{Eq15n}
	p^{BM} = \frac{95}{180}\pi^2 T^4 + \frac{T^2 \mu_B^2}{6} + \frac{\mu_B^4}{108 \, \pi^2} - B_{vac} \, .
\end{eqnarray}

Here, for the vacuum pressure $B_{vac}$, we took a conservative value $B_{vac}^\frac{1}{4} = 206$~MeV~\mbox{\cite{WongBook,Florkowski}}, which is widely accepted.

From Equation~(\ref{Eq15n}), one can find the net baryonic charge density $\rho_B$, entropy density $s$ and energy density $\epsilon$ using\- the following expressions:
\begin{eqnarray}\label{Eq16n}
	&&\rho_B^{BM} = \frac{\partial p^{BM}}{\partial \mu_B} = \frac{T^2 \mu_B}{3} + \frac{\mu_B^3}{27 \, \pi^2} \,, \\
	\label{Eq17n}
	&& s^{BM} = \frac{\partial p_{BM}}{\partial T} = \frac{19}{9}\pi^2 T^3 + \frac{T \mu_B^2}{3} \,, \\
	\label{Eq18n}
	&&\epsilon^{BM} = T s^{BM} + \mu_B\, \rho_B^{BM} - p^{BM}\, .
\end{eqnarray}

\begin{figure}[H]
\centering
\includegraphics[width=0.6\columnwidth]{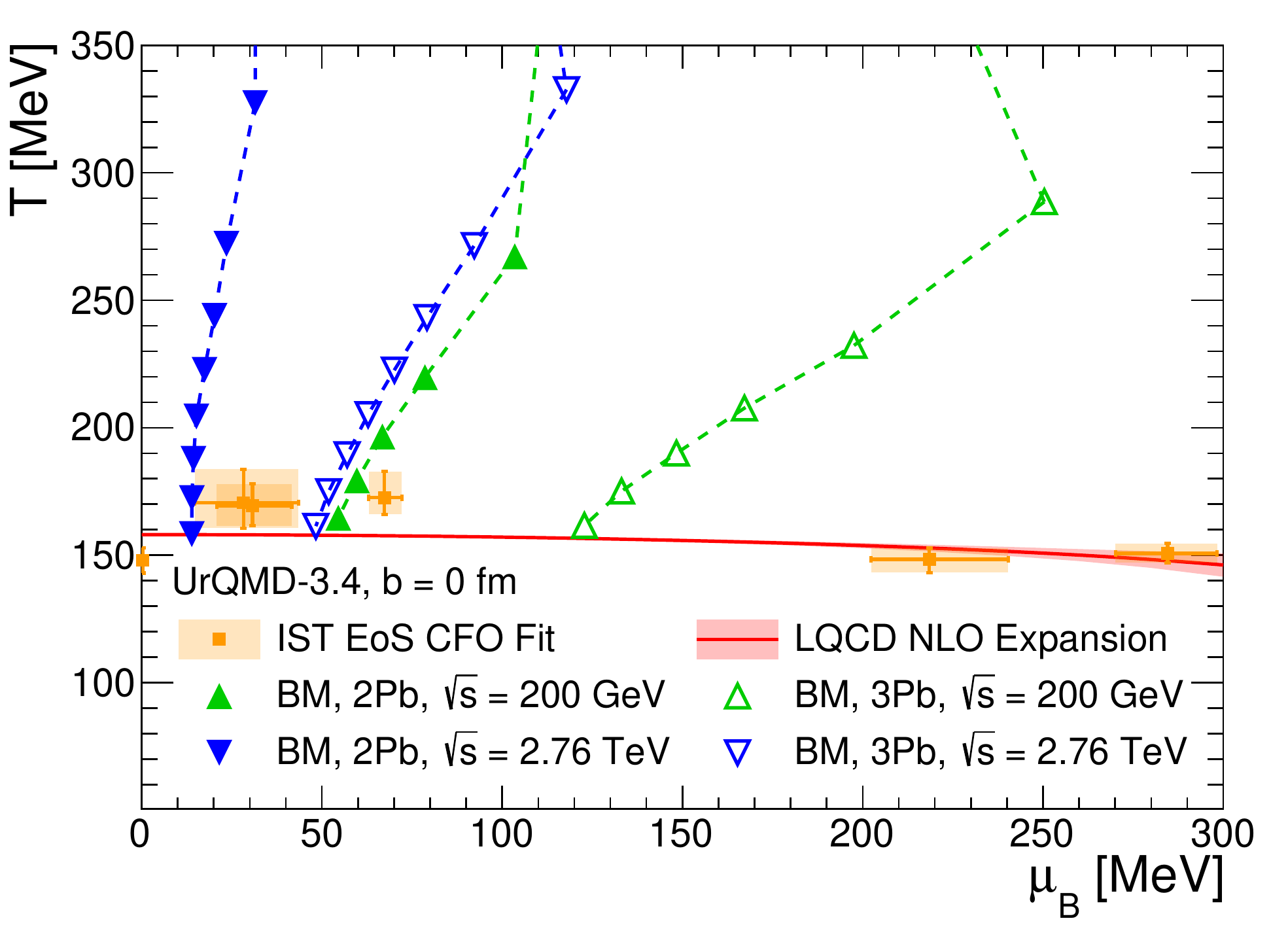}
	\caption{The evolution of central cell parameters in the $T-\mu_B$ plane obtained with the MIT bag model equation of state
~\cite{BagModelSM}. The filled symbols correspond to the Pb+Pb collisions, whereas the empty ones correspond to the Pb+Pb+Pb TNC. Collision energy $\sqrt{s_{NN}} = 200$ GeV points are shown by the triangles pointing upwards, whereas the evolution trajectories for the energy $\sqrt{s_{NN}} =2.76$ TeV are shown by the triangles pointing downwards. The topmost points correspond to the time $t-t_0 > 1$ fm. The curve of pseudo-critical temperature corresponds to a lattice QCD parameterization~\cite{lQCDparam}, whereas the crosses correspond to the parameters of chemical freeze-out in Pb+Pb collisions found in Refs.~\cite{IST1,IST2,IST3}. The collision energies of given chemical freeze-out points of Pb+Pb collisions are as follows (from left to right): $\sqrt{s_{NN}} = 2760, 200, 130, 62.4, 17.3, 12.3$ GeV~\cite{IST1,IST2,IST3}.
	}
	\label{KAB_Fig5}
\end{figure}

Using the results for $\rho_B^{BM}$ and $\epsilon^{BM}$ and equating them, respectively, to the values of net baryonic charge density and energy density obtained in the central cell for the Pb+Pb BC and  Pb+Pb+Pb TNC, we solved these equations  to find out the values of $T(t)$ and $\mu_B(t)$ at different times $t$ of evolution. Our results for $T(t)$ and $\mu_B(t)$ are depicted in Figure~\ref{KAB_Fig5} for both the TNC (empty symbols) and for the BC (filled symbols). Figure~\ref{KAB_Fig5} shows all numerical solutions obtained in this way. Appare\-ntly, they should be considered above the pseudocri\-ti\-cal temperature curve~\cite{lQCDparam} (solid curve in Figure~\ref{KAB_Fig5}) because the MIT bag model cannot be applied to the hadronic matter.

The degrees of freedom of the MIT bag model do not match the ones of UrQMD, which has a limited applicability at high temperatures. These issues require an independent test of the results on the trajectories of the central cell evolution extracted with the MIT bag model. For this, we compared the positions of the crossing of the central cell evolution trajectories and the pseudocritical temperature curve~\cite{lQCDparam} with the position of the chemical freeze-out (CFO) hypersurface in the midrapidity range~\cite{IST1,IST2,IST3}.
The chemical freeze-out parameters allow us to estimate the accuracy of our predictions obtained with the MIT bag model EoS for the baryonic chemical potential. At $\sqrt{s_{NN}} =2.76$ TeV, CFO occurs at the baryonic chemical potential $\mu_B^{CFO}\le 1$~MeV at temperature $T^{CFO}\simeq 148 \pm 2$~MeV, whereas, for $\sqrt{s_{NN}} = 200$ GeV, they are $\mu_B^{CFO}\simeq 30 \pm 3.25$~MeV and $T^{CFO}\simeq 167.3 \pm 3.9$~MeV.  From this, one can conclude that, near the CFO vicinity, the trajectories of central cell evolution shown in Figure \ref{KAB_Fig5} provide an accuracy of approximately 13~MeV for $\mu_B^{CFO}$ for Pb+Pb collisions for LHC energy and approximately $27$~MeV for RHIC top energy. In other words, the MIT bag model EoS provides a reasonable accuracy of our estimates despite its simplicity and the issues mentioned above.


In addition, it is very interesting that the  trajectory of the central cell evolution found for the Pb+Pb+Pb TNC at $\sqrt{s_{NN}} =2.76$ TeV is extremely close to the one obtained for the top RHIC energy in the Pb+Pb collisions. However, the initial net baryonic charge density of the central cell achieved at LHC energy is almost two times higher than the one achieved in the BC at RHIC.

Another interesting observation that can be drawn from Figure~\ref{KAB_Fig5} is that, near crossing the pseudocritical temperature curve, the central cell evolution trajectories demonstrate the behavior $\frac{\mu_B(t)}{T(t)} = const$, which, for the MIT bag model, automatically provides the fulfilment of condition $ \frac{s^{BM}}{\rho_B^{BM} } = const $ for each trajectory. The latter condition evidences that the central cell evolution occurs according to the hydrodynamic expansion, which also corresponds to the entropy conservation. 

\section{Conclusions and Perspectives}\label{sec.5}

In this work, we discussed the most evident aspects of the triple nuclear and hadronic collisions method. We estimated the TNC reaction rate for the p+C+p and Pb+Pb+Pb reactions under simplifying assumptions and argued that the scientific output of the TNC experiments may be useful for the location of the phase transitions predicted to exist in QCD matter and their (tri)critical endpoint(s) in the experiments. This is due to the fact that, according to our estimates, even at huge collision energies of LHC, the TNC can provide much higher net baryonic charge densities compared to the usual BC method.  
Our preliminary estimates show that, at the collision energies of beams $\sqrt{s_{NN}}$ = 10--40 GeV, the initial net baryonic charge densities can be even higher than the ones discussed here and, hence, they can lead to an even stronger enhancement of the proton and $\Lambda$-hyperon production~\cite{Vitiuk2021TNC} at these energies of collision. Thus, the TNC can be an independent and complementary source of information about the QCD matter properties.

We applied the UrQMD 3.4 model in order to study the possible effects of the transversal momentum redistribution and found them to be valuable. The conclusion about the applicability of the transport approaches such as the UrQMD to modeling processes such as the TNC, to which they were not originally designed, is an interesting result of the present work. Our simulations demonstrate that this redistribution leads to {enhancement} at a small $p_T$, which can be used for the TNC identification even for the p+C+p reactions. This effect is especially strong for protons and $\Lambda$-hyperons, whereas, in the case of pions and kaons, it can support the formation of the corresponding BEC at LHC energies.

Finally, we note that, although today, detecting TNC is a highly challenging and non-trivial experimental task, the fast technological progress in the experimental nuclear and high-energy physics allows us to hope that  the TNC rates of $10^{-3}$ s$^{-1}$ and $10^{-4}$ s$^{-1}$ can be achieved in a few years  and, hence, that the TNC physics will be accessible soon. 
In our opinion, further development of this method,  together with some practical output as a design of a new kind of fixed and jet micro-particle targets, will help the high-energy nuclear physics community to figure out how the TNC experiments can be performed. It is probable that the most  cardinal solution to produce enough TNC is a construction of a storage ring for the beam of a high intensity of ions that can be used as an inner target for the two collider beams.
  
\vspace*{4.4mm}
\noindent

\acknowledgments{Discussions with D. B. Blasch\-ke, V. Yu. Denisov,   F. Fleuert, G. Grazziani, G. Manca,  S. N. Nedelko, E. G.  Nikonov, P. Robbe,   M. Schmelling,  Ivan P. Yakimenko, D. L. Borisyuk and  all members of the LHCb IFT WG  are highly appreciated. O.V.V., K.A.B., V.M.D., S.B.C. and V.M.P.  acknowledge  support from the NAS  of Ukraine Program to enhance cooperation with CERN and JINR 
“Fundamental research on high-energy physics and nuclear physics” laun\-ched by the Section of Nuclear Physics and Energetics of NAS  of Ukraine. 
In addition, V.M.P.  acknowledges   partial  support from   LIA IDEATE  (STCU grant within the Project P9903). The simulations presented here were conducted on the JINR supercomputer Govorun. V.S. acknowledges the support from the Funda\c c\~ao para a Ci\^encia e Tecnologia (FCT) within the projects UID/04564/2021, UIDB/04564/2020, UIDP/04564/2020. The work of O.I. was supported by the Polish National Science Center under the grant No. 2019/33/BST/03059.
}


\section{Appendix}
\label{apend}

\begin{figure}[H]
\centering
{\includegraphics[width=0.665\columnwidth]{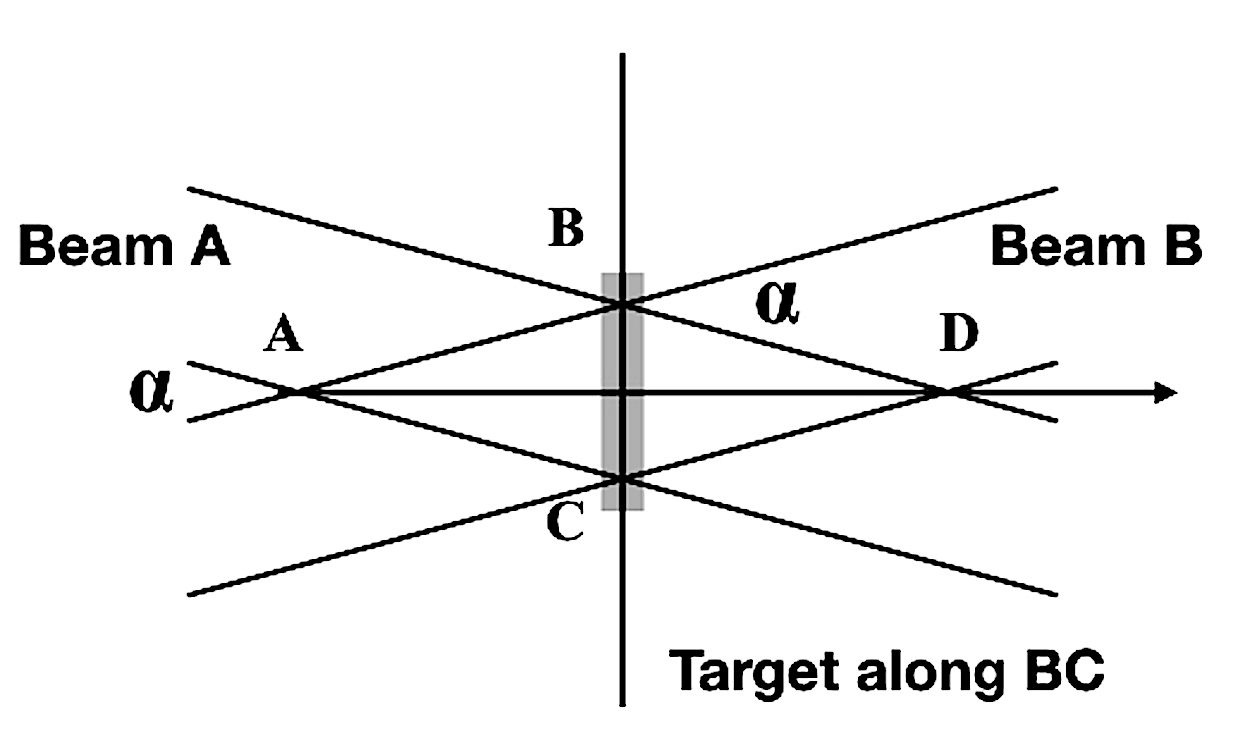}}
	\caption{\small
		Schematic picture of the interaction (intersection) volume of two identical beams A and B of a radius $R$. The angle between beams is $\alpha = 5\cdot 10^{-4}$ rad~\cite{LuminConcept} and $|BC| \simeq 2 R$. The bar interior of this volume shows a traditional arrangement of target, as discussed below.}
	\label{TNC_TargetB}
\end{figure}

For the preliminary estimates, let us consider an idealized system with a non-melting target. In addition, let us assume that the achieved interaction rates do not saturate the detector and do not affect the superconducting magnets. The nominal instantaneous luminosity of two LHC proton beams is taken as $L=10^{35}$ cm$^{-2}$ s$^{-1}$ for the collision energy $\sqrt{s_{NN}} =2.76$ TeV, which is a value expected to be achieved as a primary goal of the ATLAS and CMS Collaborations at the LHC~\cite{Colliders_MS}. According to~\cite{Colliders_MS,LuminConcept}, the number of protons per bunch $n_b=2.2\cdot 10^{11}$ 1/bunch, the number of bunches in the beam $N_{bunch}=2.8\cdot 10^3$ with the circulation frequency being $\nu=1.1245\cdot 10^4$ Hz, the nominal luminosity reduction factor $S = 0.835$ and the bunch transversal size (Gaussian standard deviation of the beam size) is $R \simeq 7~mu$m. Here, we will assume that the radii and all other properties of each beam are the same. Next, one can determine the longitudinal size $|AD|$ of the interaction zone of the beams A and B (see Figure \ref{TNC_TargetB}) with the help of the above mentioned estimates for $R$. One can approximately obtain
\begin{eqnarray}\label{Eq6}
    |BC|\hspace*{-0.33mm} &=&\hspace*{-0.33mm}\frac{2R}{\cos(\alpha/2)} \hspace*{-0.33mm} \simeq\hspace*{-0.33mm} 2R ,\\
	|AB|\hspace*{-0.33mm} &=& \hspace*{-0.33mm} |BD|\hspace*{-0.33mm} =\hspace*{-0.33mm} \frac{2R}{\sin(\alpha)}\hspace*{-0.33mm} \simeq\hspace*{-0.33mm} \frac{2R}{\alpha}, \\
	|AD| &=& \frac{2R}{\sin(\alpha/2)} \simeq \frac{4R}{\alpha} = 2 |AB| 
	\,,\quad 
	\label{Eq7}
\end{eqnarray}
since the standard angle of the beam intersection for LHC $\alpha = 5\cdot 10^{-4}$ rad is small~\cite{LuminConcept}. Evaluating these expressions, one obtains $|AD| \simeq 5.6 $ cm and $|AB| =|AC| = |CD| \simeq 2.8 $ cm.
\vspace{-12pt}
For completeness, first, let us consider a carbon target with  the density of $\rho_C = 2.2$ g $\cdot$ cm$^{-3}$ as shown in Figure \ref{TNC_TargetB}, which  is located perpendicular to the interaction volume of two colliding beams, with the geometrical thickness $l_g $  being of a few $\mu$m,
which we measure along the line   $|AD|$. The total number of protons emitted from both beams hitting the target per one second is $\frac{d N_p}{d t} \simeq 2 n_b N_{bunch} \nu = 1.4\cdot 10^{19}$ s$^{-1}$.

Since the carbon molar mass ${\cal M}$ is $12$ g/mole and $N_A$ denotes Avogadro's number, then, for the target of geometrical thickness  $l_g= 3.32\, \mu$m  one finds that the  thickness of the target is $\tau_C \simeq l_g \, \rho_C N_A / {\cal M} \sim 3.65 \cdot 10^{19}$ atoms$\cdot$cm$^{-2}$.
Hence,  the number of binary p+C collisions per second produced by two proton beams is
\begin{equation}\label{Eq8}
	\frac{d N_{p+C}}{d t} \simeq  \frac{d N_p}{d t} \, \tau_C \, \sigma_{p+C} \simeq 5.1 \cdot 10^{13} ~ {\rm s}^{-1},
\end{equation} 
\noindent
where, for typical LHC energies, the inelastic cross-section of the p+C collision is taken to be $\sigma_{p+C} \simeq 100 $ mb $ \simeq 10^{-25}$ cm$^{-2}$. One should note that our assumption on the chosen inelastic cross-section $\sigma_{p+C} \simeq 100 $ mb is consistent with the geometrical estimate for the cross-section
\begin{equation}\label{Eq9n}
	\sigma_{A+T} \simeq \sigma_{p+C} \left[ \frac{A^\frac{1}{3}+T^\frac{1}{3}}{1^\frac{1}{3}+12^\frac{1}{3}} \right]^2
	\,, 
\end{equation}
which is a simplified version of the geometrical formulas~\cite{Geometrical_CS} relating the p+p, p+A and A+B collisions. Indeed, using Equation~(\ref{Eq9n}) for $A=1$ and $T=1$ and taking the inelastic p+p cross-section as $\sigma_{p+p} \simeq 35$ mb at the collision energy $\sqrt{s_{pp}} \simeq 63$ GeV~\cite{PDG},  one can obtain $\sigma_{p+C} \simeq 2.7 \sigma_{p+p} \simeq 94.5$ mb, which approximately corresponds to the proton beam energy for a fixed target experiment.

\begin{figure}[H]
\centering
	{\includegraphics[width=0.56\columnwidth]{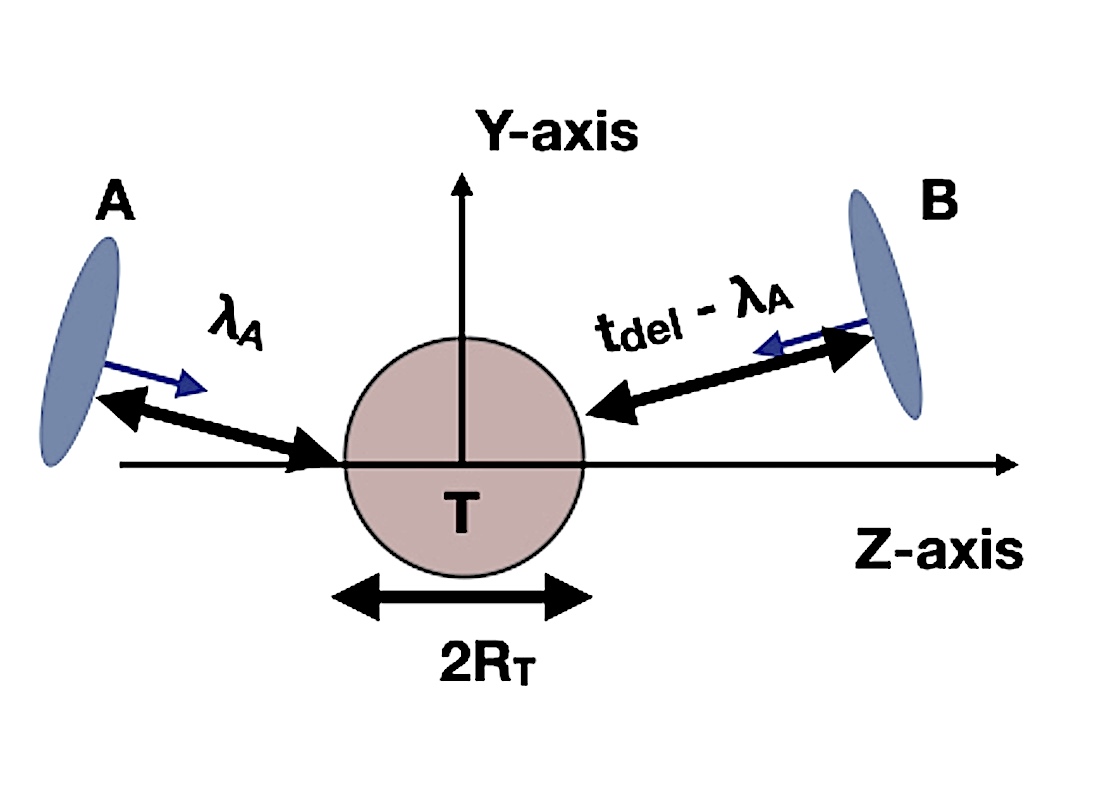}}
	\caption{\small
		Scheme of the interaction process of two colliding nuclei A and B with the nucleus of the target T with the geometrical radii $R_A$, $R_B$ and $R_T$, respectively. The interaction volume $V^{int}_{T\in A+B} $ is defined by the distance between the surfaces of nuclei A and B, which is $2 R_T + t_{del}$, and by the maximal geometrical size of nuclei A and T. For such a configuration of nuclei, the TNC will occur with a probability equal to $1$. The mean value for $V^{int}_{T\in A+B}$ is provided by averaging over the distance $\lambda_A$ between the surfaces of a nucleus A and the target T. Similar averaging over the nucleus B position should be performed as well.
	}
	\label{TNC_Tdel}
\end{figure}

Assuming that the protons from the beams A and B are approaching the target nucleus T and each other as shown in Figure~\ref{TNC_Tdel},  let us consider the p+C+p TNC as a p+p collision occurring inside the carbon target with the time delay $t_{del} \le 10$ fm and denote the distances of proton A and B from the surface of nucleus T as $ \lambda_A$ and $t_{del} - \lambda_A$, respectively. According to our simulations, such a delay will not produce a significant difference in the final state in comparison with the simultaneous TNC. For the given time delay, one can find the mean distance between the particles of beams A and B to be $D_{AB} = 2 R_T + t_{del} /2$ by varying the distance $ \lambda_A$ in the limits from 0 to $t_{del} $ with equal probability. 
Apparently,  one obtains the same  mean distance $D_{AB} = 2 R_T + t_{del} /2$ between the particles of beams A and B if one averages the position  of the nucleus B  under the same conditions. 
\vspace{-10pt}

From Figure~\ref{TNC_Tdel}, it is clear that the p+C+p TNC with the time delay $t_{del} \le 10$ fm can occur in the case of some non-zero number of target nuclei present inside the interaction volume of three nuclei $V^{int}_{T\in A+B} \simeq \pi \left[ \left(\max\{R_A; R_T\}\right)^2 + \left(\max\{R_B; R_T\}\right)^2\right] \cdot D_{AB}$. For the given density of target $\rho_T$, one can obtain the number of target nuclei entering the interacting volume $V^{int}_{T\in A+B}$ as $N^{int}_{T\in A+B} = \rho_T V^{int}_{T\in A+B}$. Hence, the expected rate of the TNC is given as
\begin{equation}\label{Eq10n}
	 \frac{d N_{A+T+B}}{d t} \simeq L \cdot \sigma_{A+B} \cdot \rho_T V^{int}_{T\in A+B} \cdot \frac{V_T^{irrad}}{V_{beams}}
	\,,
\end{equation}
where the first two multipliers on the right-hand side correspond to the standard rate of binary collisions A+B. It is necessary to stress that Equation~(\ref{Eq10n}) provides the TNC rate for the arrangement of the target shown in Figure~\ref{TNC_TargetB}, i.e., with  the arrangement along the  line $|BC|$ in Figure~\ref{TNC_TargetB}. 
The volume of the target irradiated  by the beams is $V_T^{irrad} \simeq \pi R^2 l_g$, whereas the interaction volume of two beams is $V_{beams} \simeq  \pi R^2 |AD| \simeq 4 \pi R^3 [\sin(\alpha)]^{-1}$.  The reduction factor $ \frac{V_T^{irrad}}
{V_{beams}}$ in Equation~(\ref{Eq10n}) accounts for the fact that only the part of binary reactions among the beam particles 
occurs inside the target. To simplify our estimates, we considered the interaction volume of  beams  to be a cylinder of radius $R$ and height $ |AD| $. 
For the geometrical thickness of target $l_g= 3.34\, \mu$m, one obtains the following result
\begin{equation}\label{Eq10nB}
	\frac{V_T^{irrad}}{V_{beams}} \simeq \frac{l_g \sin(\alpha)}{4 R} \simeq 0.57 \cdot 10^{-5} 
	 \,.~
\end{equation}

Taking the radius of the nucleus of A nucleons as $R_A \simeq 1.25 \cdot A^\frac{1}{3}$ fm~\cite{Krane}, 
for the traditional arrangement of the target, one can obtain the p+C+p TNC rate as
	\begin{equation}\label{Eq11n}
		\frac{d N_{p+C+p}}{d t} \simeq 4.8 \cdot 10^{-9} \frac{1}{s},
	\end{equation}
where the inelastic cross-section of  p+p BC was taken as $ \sigma_{p+p} = 35 $ mb~\cite{PDG} and $ \rho_C V^{int}_{C\in p+p} \simeq 6 \cdot 10^{-14}$ nucl. 

Although the result of Equation~(\ref{Eq11n}) {corresponds to a low collision rate, and, consequently an insufficient number of events per unit of time,} we note that, in a few years, it can be greatly improved if the new types of targets will be used (see the discussion below). 

For instance, the simplest way to increase the TNC reaction rate is to arrange the target in such a non-traditional way
that it completely fills  the interaction volume of two colliding beams (it has to be arranged along the line  $|AD|$ in
Figure~\ref{TNC_TargetB}). In this case, for its
geometrical thickness  $l_g = |BC| \simeq 3.34$ $\mu$m,  the  factor 
$\frac{V_T^{irrad}}{V_{beams}}=1$  does not suppresses the reaction rate and,  instead of Equation~(\ref{Eq11n}),
one finds
 \begin{eqnarray}\label{Eq11nB}
&& \frac{d N_{p+C+p}}{d t} \simeq  2 \hspace*{-0.33mm}\cdot \hspace*{-0.33mm} 10^{-4}  
 \, \frac{1}{s}
	 \,,~
\end{eqnarray}
which is essentially higher compared to the result of Equation~(\ref{Eq11n}).
This result is  more optimistic since the passage of beam particles in such a case is $\lambda_p \simeq |AC| \simeq 2.8 $ cm
and, hence, the target thickness becomes $\tau_C \simeq \lambda_p \, \rho_C N_A / {\cal M} \sim 3.1 \cdot 10^{23}$ atoms$\cdot$cm$^{-2}$.

Now, let us consider the general type of  the A+T+B collisions. Geometrically, the inelastic cross-section of A+T collisions can be approximately expressed in terms of the p+C cross-section according to Equation~(\ref{Eq9n}).
One should note that this is a simplified version of the geo\-metrical cross-sections relating the p+p, p+A and A+B collisions given by Ref.~\cite{Geometrical_CS}. Assuming  the Pb beams luminosity value $L \simeq 10^{28}$ cm$^{-2}$ s$^{-1}$ will allow us to obtain the rate of the Pb+Pb+Pb TNC from Equations (\ref{Eq9n}) and (\ref{Eq10n}) 
 as
  \begin{eqnarray}\label{Eq13n}
&& \frac{d N_{3Pb}}{d t} \simeq \hspace*{-0.33mm}10^{28}\hspace*{-0.33mm} \cdot \hspace*{-0.33mm} 1.3 \hspace*{-0.33mm}\cdot \hspace*{-0.33mm} 10^{-24} \hspace*{-0.33mm} \cdot \hspace*{-0.33mm} 2.25  \hspace*{-0.33mm} \cdot  \hspace*{-0.33mm} 10^{-13} \hspace*{-0.33mm} \cdot \hspace*{-0.33mm} \frac{V_T^{irrad}}{V_{beams}} \simeq  \\
&& \simeq  \left\{ \begin{array}{ll}
      2.4 \cdot 10^{-11} ~ \frac{1}{s}    & \mbox{for traditional arrangement,}\\
          & \\  \label{Eq13nB}
        3.4 \cdot 10^{-7} ~ \frac{1}{s} & \mbox{for non-traditional arrangement,}
        \end{array} \right.  
\end{eqnarray}
  where the 
 value of the geometrical thickness of the lead target was taken as $l=3.34 \, \mu$m, $ \sigma_{Pb+Pb} \simeq 1.3\cdot 10^{-24} $ cm$^2$ and the density of lead $\rho_{Pb} \simeq 11.4$ g$\cdot$cm$^{-3}$, whereas  the ratio of volumes $\frac{V_T^{irrad}}{V_{beams}} $ was taken from Equation~(\ref{Eq10nB}) for the traditional arrangement of the target (as shown in Figure~\ref{TNC_TargetB}) and  $\frac{V_T^{irrad}}{V_{beams}} =1 $ for the non-traditional method of target arrangement
 (along the line $|AD|$ in
Figure~\ref{TNC_TargetB}).

The abovementioned estimates were performed for the lower limit of the TNC rates. Estimating the interaction volume of three nuclei $V^{int}_{T\in A+B} $ with the help of the cross-sections of the code EPOS~\cite{EPOS} instead of the geometrical cross-sections, one should multiply the respective rates  from  Equations~(\ref{Eq11n}), (\ref{Eq11nB}) and (\ref{Eq13nB}) by a factor of four.

We would like to emphasize that the TNC method allows one to study the interaction of nuclei with the hot fireball formed by the collision of another pair of nucleus among other phenomena, which is a completely novel approach and has not yet been studied.

Since the TNC is a novel idea, there are  more than enough unsettled problems that
cannot be fully discussed in this work. In particular, the detection of the TNC is of primary importance. In the Section \ref{sec.3}, this question is 
discussed for both p+C+p and Pb+Pb+Pb TNC.
In addition, it seems that   an increase in the intersection angle between the beams to the value of $1/30$--$1/15$ rad can essentially simplify the detection of TNC. 

The other  essential issue is the question of energy deposition to the target by two bombarding beams. Let us show that this is possibly an issue for the  non-traditional arrangement of the target, as suggested above. Using the Particle Data Group results~\cite{PDG}  on the mean rate of energy loss (or stopping power) 
$\left| \frac{d E}{d x} (T) \right|_A$ of particles of type A on target T, one can estimate the energy deposition of two beams on the target. For the carbon target  arranged non-traditionally, which is bombarded  by two proton beams, one can obtain
\begin{eqnarray}\label{EqA14n}
	&& \hspace*{-2.2mm}\frac{d E_{p+C+p}}{d t} \simeq \frac{d N_p}{d t} \cdot \left| \frac{d E (C)}{d x} \right|_p \rho_C \lambda_p
	\simeq ~ \nonumber \\
	~
	\label{EqA15n}
	&&  \hspace*{-2.2mm}\simeq 1.4\cdot 10^{19} \frac{1}{s} \cdot 2.2 \frac{\rm~MeV\cdot cm^2}{g} 2.2 \frac{g}{\rm cm^3} \cdot 2.8\, {\rm cm} \simeq ~  \nonumber \\
	\label{EqA16n}
	&&  \hspace*{-2.2mm} \simeq 3 \cdot 10^7 \frac{J}{s} ,~
\end{eqnarray}
where, in Equation~(\ref{EqA14n}),  we substituted our above estimates for $\frac{d N_p}{d t} \simeq 1.4\cdot 10^{19} \frac{1}{s} $ and $\lambda_p =2.8$ cm and the value $\left| \frac{d E (C)}{d x} \right|_p \simeq 2.2 \frac{\rm~MeV\cdot cm^2}{g} $ taken  from the Particle Data Group~\cite{PDG} for the energy loss of protons in the carbon target. Apparently, the obtained value of $\frac{d E_{p+C+p}}{d t}$ is extremely large and, hence, 
the non-traditional arrangement of the fixed target is not suited for the TNC, as it leads to too high energy deposit to the target.

Note that the high-energy deposition given by Equation~(\ref{EqA16n}), obtained for the non-traditional  arrangement of the target,  can be reduced by the factor $ F = \frac{ 2\lambda_p 2 R }{2\lambda_p 2 R + \lambda_p \cdot v_T \cdot 1 s }$ in the case of the target surface moving with the speed $v_T$. For the linear speed of such a surface of approximately $33$ m/s, the reduction factor is $F \simeq 10^{-7}$ and, therefore, the rota\-ting carbon target will be able to withstand irradiation with proton beams without the risk of explosion. 

For the carbon target irradiated by two proton beams, one can find out that the rate of the increase in target temperature ${\cal T}_C$ is rather high, even for small values of the reduction factor $F$. Indeed, the mass of the target that experiences instantaneous irradiation is ${\cal M}_T \simeq \lambda_p (2R)^2 \cdot \rho_C \simeq 1.2\cdot 10^{-5}$ g. Neglecting the possible heat loss, one can obtain the carbon target temperate increase rate as
\begin{equation}\label{EqA17n}
	\frac{d {\cal T}_C}{d t} \simeq \frac{d E_{p+C+p}}{d t} \cdot F \cdot \left[ C_c \cdot {\cal M}_T \right]^{-1} \simeq 1.5 \cdot 10^{5}\, \frac{K}{s}
	\,,
\end{equation}
where we used the heat capacity of the carbon target (gra\-phite) as $C_c \simeq 1.7$ J/g/K, which is valid for ${\cal T}_C$ above $700$ K. Thus, the irradiated fragments of the thin target will evaporate quickly and, hence, the target should be made restorable.

With the help of the Bethe--Bloch equation~\cite{PDG} for the stopping power, one can approximately find  out that the energy deposition $\frac{d E_{3Pb}}{d t} $ is at least a hundred times larger than the result of Equation~(\ref{EqA14n}) for the p+C+p TNC.


\end{document}